\documentclass[12pt]{article}
\usepackage{amsmath,amssymb,bbold,hyperref,graphicx}

\newcommand{\be}{\begin{equation}}
\newcommand{\ee}{\end{equation}}
\newcommand{\bea}{\begin{eqnarray}}
\newcommand{\eea}{\end{eqnarray}}
\newcommand{\beas}{\begin{eqnarray*}}
\newcommand{\eeas}{\end{eqnarray*}}

\newcommand{\UU}{\rotatebox{90}{\mbox{${\textstyle \sf U}$}} \hspace{2mm} \raisebox{2.8mm}{\rotatebox{-90}{\mbox{${\textstyle \sf U}$}}}}
\newcommand{\rightU}{\raisebox{2.5mm}{\rotatebox{-90}{\mbox{${\textstyle \sf U}$}}}\,\,}
\newcommand{\leftU}{\raisebox{-0.2mm}{\rotatebox{90}{\mbox{${\textstyle \sf U}$}}}\,\,}
\newcommand{\Hext}{\overset{\leftrightarrow}{H}\rlap{\phantom{H}}}
\newcommand{\Hextvac}{\overset{\leftrightarrow}{H}\rlap{\phantom{H}}^{(0)}}
\def\identity{{\mathbb{1}}}

\begin{document}
\begin{titlepage}

\begin{center}

{\Large Light-ray moments as endpoint contributions to modular Hamiltonians}

\vspace{12mm}

\renewcommand\thefootnote{\mbox{$\fnsymbol{footnote}$}}
Daniel Kabat${}^{1,2}$\footnote{daniel.kabat@lehman.cuny.edu},
Gilad Lifschytz${}^{3}$\footnote{giladl@research.haifa.ac.il},
Phuc Nguyen${}^{1,3}$\footnote{phuc.nguyen@lehman.cuny.edu},
Debajyoti Sarkar${}^{4}$\footnote{dsarkar@iiti.ac.in}

\vspace{6mm}

${}^1${\small \sl Department of Physics and Astronomy} \\
{\small \sl Lehman College, City University of New York} \\
{\small \sl 250 Bedford Park Blvd.\ W, Bronx NY 10468, USA}

\vspace{2mm}

${}^2${\small \sl Graduate School and University Center, City University of New York} \\
{\small \sl  365 Fifth Avenue, New York NY 10016, USA}

\vspace{2mm}

${}^3${\small \sl Department of Mathematics and} \\
{\small \sl Haifa Research Center for Theoretical Physics and Astrophysics} \\
{\small \sl University of Haifa, Haifa 3498838, Israel}

\vspace{2mm}

${}^4${\small \sl Department of Physics} \\
{\small \sl Indian Institute of Technology Indore} \\
{\small \sl Khandwa Road 453552 Indore, India}

\end{center}

\vspace{12mm}

\noindent
We consider excited states in a CFT, obtained by applying a weak unitary perturbation to the vacuum.  The perturbation is generated by the
integral of a local operator $J^{(n)}$ of modular weight $n$ over a spacelike surface passing through $x = 0$.
For $\vert n \vert \geq 2$ the modular Hamiltonian associated with a division of space at $x = 0$ picks up an endpoint contribution,
sensitive to the details of the perturbation (including the shape of the spacelike surface) at $x = 0$.  The endpoint contribution is a sum of light-ray moments of the perturbing operator $J^{(n)}$ and its descendants.
For perturbations on null planes only moments of $J^{(n)}$ itself contribute.

\end{titlepage}
\setcounter{footnote}{0}
\renewcommand\thefootnote{\mbox{\arabic{footnote}}}

\hrule
\tableofcontents
\bigskip
\hrule

\addtolength{\parskip}{8pt}

\section{Introduction\label{sect:intro}}
The modular operator $\Delta$, and the associated extended modular Hamiltonian $- \log \Delta$, are fundamental objects in the study of von Neumann algebras.
They provide powerful tools for working with spatial subregions of a quantum field theory \cite{Haag:1992hx,Borchers:2000pv,Witten:2018lha}.  Unfortunately explicit expressions for modular Hamiltonians are hard to come by.
For the Poincar\'e vacuum and a division into half-spaces the modular Hamiltonian can be identified with a Lorentz generator \cite{Bisognano:1976za}, while for a CFT in its
ground state and a division into spherical regions the modular Hamiltonian can be identified with a linear combination of conformal generators \cite{Hislop:1981uh}.  A few more examples are available in the literature \cite{Cardy:2016fqc}.  But not
much is explicitly known about the modular Hamiltonian for a generic excited state.

In this paper we take a small step toward a more generic situation, by constructing the modular Hamiltonian for a state in a CFT that is obtained by applying a weak unitary
perturbation to the vacuum.  At $t = 0$ we make a division into half-spaces and compute the modular Hamiltonian to first order in the strength of the perturbation.
The surprise is that we find an endpoint contribution to the modular Hamiltonian, built from light-ray moments of the perturbing operator and its descendants, that is sensitive
to the details of the perturbation at $x = 0$.

In more detail, consider a Lorentzian field theory and a state $\vert \psi \rangle = e^{-i \epsilon G} \vert 0 \rangle$ created from the vacuum by a unitary operator.  We will
consider more general perturbations below, but for now suppose $G$ acts along a one-dimensional spatial segment at $t = 0$.
It will be convenient to work in light-front coordinates $x^\pm = t \pm x$, so we will write $G$ as
\be
\label{Gdef}
G = \int_{-b}^a dx f(x) J^{(n)}(x^+= x,x^- = -x,0)
\ee
Here $f(x)$ is a real function and $J^{(n)}(x^+,x^-,{\bf x}_\perp)$ is a local operator with modular weight $n$, meaning that under a Lorentz boost
\be
\label{boost}
J^{(n)}(x) \rightarrow e^{ns} J^{(n)}(e^s x^+, e^{-s} x^-, {\bf x}_\perp)
\ee
We'll only consider operators with integer $n$; for example $T_{++}$ has $n = 2$.
The limits on the $x$ integral ensure that the perturbation has support on a finite interval $-b < x < a$.  For our purposes it won't matter whether $f$ vanishes at the ends of the interval.  We'll work to first order in the expansion parameter $\epsilon$.

At $t = 0$ we divide space into $A \cup \bar{A}$, with $A = \lbrace x > 0 \rbrace$ and $\bar{A} = \lbrace x < 0 \rbrace$.  We'll refer to $x = 0$ as the endpoint.\footnote{
If either $a = 0$ or $b = 0$ then $G$ just touches the endpoint.  This will make the results ambiguous, analogous to the ambiguity in defining $\int_0^\infty dx \, \delta(x) f(x)$.
To avoid this we'll assume throughout that $a$ and $b$ are non-zero.}  It would seem that since $G$ is the integral over $x$ of a local operator
it can be decomposed into
\be
\label{G}
G = G_A \otimes \identity_{\bar{A}} + \identity_A \otimes G_{\bar{A}}
\ee
where
\be
\label{GA}
G_A = \int_0^a dx f(x) J^{(n)}(x^+ = x,x^- = -x,0)
\ee
just acts on region $A$ while
\be
\label{GAbar}
G_{\bar{A}} = \int_{-b}^0 dx f(x) J^{(n)}(x^+ = x,x^- = -x,0)
\ee
just acts on region $\bar{A}$.  As we proceed we'll see that the decomposition (\ref{G}) is more subtle than it appears.  It can be used but must be treated with care.  This seems to be related to the fact
that the Hilbert space of a field theory can't be factored into a tensor product ${\cal H}_A \otimes {\cal H}_{\bar{A}}$.

We're interested in computing the first-order change in the modular Hamiltonian for region $A$, denoted $\delta H_A$.  For this we use the S\'arosi--Ugajin formula \cite{Sarosi:2017rsq}
or the related expansion developed in \cite{Lashkari:2018tjh}.  As shown in \cite{Kabat:2020oic}, for a perturbation of the form (\ref{G}) this leads to\footnote{A note of caution: in \cite{Kabat:2020oic} $G$ was decomposed
into a sum of tensor products of operators on $A$ and $\bar{A}$.
In the present work we don't invoke such a decomposition.  Instead we will only use (\ref{G}), which is the special case where one of the operators is the identity.
This simplifies the calculation but slightly changes the meaning of $G_A$ and $G_{\bar{A}}$.}
\be
\label{SUformula}
\delta H_A = {i \epsilon \over 2} \int_{-\infty}^\infty {ds \over 1 + \cosh s} \left((G_A \big\vert_{s - i \pi} + \widetilde{G_{\bar{A}}} \big\vert_s) - (G_A \big\vert_{s + i\pi}
+ \widetilde{G_{\bar{A}}} \big\vert_s) \right)
\ee
Here ${\cal O}\vert_s$ denotes vacuum modular flow, ${\cal O}\vert_s = \left(\Delta^{(0)}\right)^{-is/2\pi} {\cal O} \left(\Delta^{(0)}\right)^{is/2\pi}$, and $\widetilde{\cal O}$ denotes CPT conjugation.
We present a path-integral derivation of this result in appendix \ref{appendix:SUformula}.

Our goal in the rest of this paper is to extract an explicit result for $\delta H_A$ from the rather formal expression (\ref{SUformula}).
The result for $\delta H_A$ can be combined with an analogous result for the complementary region to obtain the first-order
change in the extended or total modular Hamiltonian $\delta \Hext = \delta H_A - \delta H_{\bar{A}}$.  As in \cite{Kabat:2020oic} we find that $\delta \Hext$
is a sum of commutator and endpoint contributions.
\be
\label{deltaHext}
\delta \Hext = \delta \Hext^{\rm commutator} + \delta \Hext^{\rm endpoint}
\ee
The commutator contribution captures the naive expectation that $\Hext$ is related to the vacuum modular
Hamiltonian $\Hext^{(0)}$ by an infinitesimal unitary transformation.\footnote{\label{factorize}If the unitary transformation factorizes, $U = U_A \otimes U_{\bar{A}}$ so that $\vert \psi \rangle = \big(U_A \otimes U_{\bar{A}}\big) \vert 0 \rangle$,
then the modular operator (formally given by $\Delta = \rho_A \otimes \rho_{\bar{A}}^{-1}$) is related to the vacuum modular operator $\Delta^{(0)}$ by $\Delta = U \Delta^{(0)} U^\dagger$.  For $\Hext = - \log \Delta$ this leads to (\ref{NaiveCommutator}).}
\be
\label{NaiveCommutator}
\delta \Hext^{\rm commutator} = - i \epsilon [G,\Hext^{(0)}]
\ee
The endpoint contribution gives a correction to this result, arising from the fact that in general the unitary transformation
doesn't factorize between $A$ and $\bar{A}$.
Extending previous results \cite{Kabat:2020oic} we find that the endpoint contribution involves operators we will refer to as light-ray moments of $J^{(n)}$ and its
descendants.  For planar surfaces the general result will be given in (\ref{TransverseG}) - (\ref{TransverseNegative}), and for perturbations with positive modular weight on
a curved surface the general result will be given in (\ref{endpoint4}).  As a preview, for the special case of the perturbation (\ref{Gdef}) that acts on a spatial segment at $t = 0$ we find ($f^{(k)} = \partial_x^k f(x)$)
\begin{itemize}
\item
The endpoint contribution vanishes for modular weights $n = -1,0,1$.  For an intuitive explanation of this fact see appendix D of \cite{Kabat:2020oic}: it requires a perturbation of modular weight $\vert n \vert \geq 2$,
such as the $T_{++}$ and $T_{--}$ components of the stress tensor, to move degrees of freedom between $A$ and $\bar{A}$.
\item
For modular weights $n = 2,3,4,\ldots$ the endpoint contribution is a sum of light-ray moments of $J^{(n)}$ and its descendants on the $x^+$ horizon.
\be
\label{positive}
\delta \Hext^{\rm endpoint} = - 2 \pi \epsilon \sum_{k=0}^{n-2} {1 \over k!} f^{(k)}(0) \sum_{l = 0}^{\left\lfloor {n - 2 - k \over 2} \right\rfloor} {(-1)^l \over l!} (n - k - 2l - 1)
\int_{-\infty}^\infty dx^+ (x^+)^{k+l} \partial_-^l J^{(n)}(x^+,0,0)
\ee
\item
For modular weights $n = -2,-3,-4,\ldots$ the endpoint contribution is a sum of light-ray moments of $J^{(n)}$ and its descendants on the $x^-$ horizon.
\be
\label{negative}
\delta \Hext^{\rm endpoint} = + 2 \pi \epsilon \sum_{k=0}^{\vert n \vert-2} {(-1)^k \over k!} f^{(k)}(0) \sum_{l = 0}^{\left\lfloor {\vert n \vert - 2 - k \over 2} \right\rfloor}
{(-1)^l \over l!} (\vert n \vert-k-2l-1) \int_{-\infty}^\infty dx^- (x^-)^{k+l} \partial_+^l J^{(n)}(0,x^-,0)
\ee
\end{itemize}
Here $\lfloor \cdot \rfloor$ denotes the floor function.  The results (\ref{positive}), (\ref{negative}) are exchanged by parity, as we discuss in section \ref{sect:generalize}.

Note that our previous calculations \cite{Kabat:2020oic} were for perturbations that acted on a null segment, and were incomplete in that they only detected operators
corresponding to the $k = 0$ term in these sums.  In the present work we will treat general spacelike surfaces, including a discussion of the null limit, and we will find the full
set of light-ray moments that appear in endpoint contributions.  We should point out that in two dimensions general stress tensor perturbations have been studied by Das and Ezhuthachan
\cite{Das:2018ojl}, and as noted in \cite{Kabat:2020oic} we recover their results for weak perturbations and divisions into half-spaces.  For other studies of modular Hamiltonians for excited
states see \cite{Allais:2014ata,Lashkari:2015dia,Faulkner:2015csl,Sarosi:2016oks,Faulkner:2016mzt,Casini:2017roe,Lewkowycz:2018sgn,Longo:2018obd,Lashkari:2018oke,deBoer:2019uem,Rosso:2019txh,Balakrishnan:2020lbp,Arias:2020qpg,Rosso:2020cub}.
We should also point out that light-ray operators are well-known in the literature \cite{Kravchuk:2018htv}.  For example the average null energy operator \cite{Faulkner:2016mzt} is the zeroth moment of the stress tensor
while higher moments have been studied in \cite{Casini:2017roe,Cordova:2018ygx,Kologlu:2019bco,Huang:2020ycs,Belin:2020lsr}.

An outline of this paper is as follows.  In section \ref{sect:preliminaries} we discuss preliminary points needed to set up the calculation.  In section \ref{sect:GAcorrelators} we discuss
properties of correlators of $G_A$ and $\widetilde{G_{\bar{A}}}$ that play a role in the calculation.  In section \ref{sect:deltaHAcorrelators} we obtain a general formula for correlators of $\delta H_A$ and
in section \ref{sect:extract} we extract an operator expression for $\delta H_A$ from these correlators.  Section \ref{sect:generalize} generalizes and extends the results and we conclude in
section \ref{sect:conclusions}.
We present a path integral derivation of the
S\'arosi--Ugajin formula in appendix \ref{appendix:SUformula} and some explicit CFT correlators in appendices \ref{appendix:correlators} and \ref{appendix:yperp}.

\section{Preliminaries\label{sect:preliminaries}}
Our goal is to use the S\'arosi--Ugajin formula (\ref{SUformula}) to obtain an explicit expression for $\delta H_A$.  However we need to address a number of preliminary points
before we can make use of (\ref{SUformula}).

\begin{figure}
\centerline{\includegraphics{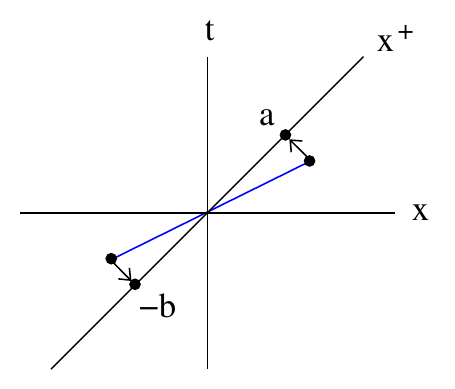}}
\caption{The perturbation acts on the spacelike segment shown in blue.  As $\theta \rightarrow 0^+$ the segment approaches a piece of the $x^+$ horizon.\label{fig:regulate2}}
\end{figure}

In what follows it will be useful (and not difficult) to generalize the perturbation so that it acts on a tilted space-like segment.  We do this by introducing a tilt parameter $\theta$ and setting
$x^- = - \theta x^+$ as shown in Fig.\ \ref{fig:regulate2}.  That is, we consider the perturbation
\be
\label{Gthetadef}
G = \int_{-b}^a dx^+ f(x^+) J^{(n)}(x^+,-\theta x^+,0)
\ee
In this formula note that $x^+$ is just an integration variable used to parametrize the segment; we are labeling points along the segment by the value of their $x^+$
coordinate.  The purely spatial perturbation (\ref{Gdef}) corresponds to setting $\theta = 1$.  Perturbations along the $x^+$ horizon can be studied by sending $\theta \rightarrow 0^+$
while perturbations along the $x^-$ horizon can be studied, a bit more awkwardly, by sending $\theta \rightarrow \infty$.  We will consider these limits in section \ref{sect:generalize},
where we will also consider the generalization to curved surfaces.

\noindent
Assuming the segment is spacelike, $0 < \theta < \infty$, it would appear we can decompose $G$ into operators that act on $A$ and $\bar{A}$ as
\be
\label{Gtheta}
G = G_A \otimes \identity_{\bar{A}} + \identity_A \otimes G_{\bar{A}}
\ee
where
\begin{eqnarray}
\label{GAtheta}
&& G_A = \int_0^a dx^+ f(x^+) J^{(n)}(x^+,-\theta x^+,0) \\
\label{GAbartheta}
&& G_{\bar{A}} = \int_{-b}^0 dx^+ f(x^+) J^{(n)}(x^+,-\theta x^+,0)
\end{eqnarray}
One might worry this leaves the status of the endpoint at $x = 0$ (or equivalently at $x^+ = 0$) ambiguous: does it belong to $A$
or $\bar{A}$?  As we'll see it's possible to keep track of the endpoint contribution in a  consistent manner, by working inside correlation
functions and paying careful attention to singularities.  We'll never have to address the question of whether $x = 0$ belongs to $A$ or $\bar{A}$.

In attempting to use the decomposition (\ref{Gtheta}) -- (\ref{GAbartheta}) with the S\'arosi--Ugajin formula (\ref{SUformula}) there are two practical concerns we face.
First, the S\'arosi--Ugajin formula involves an integral over modular time, and there are potential divergences at early and late modular times.
Second, the formula involves complex modular flow, and we need a prescription for making sense of that.

To deal with potential divergences in the integral over modular time we introduce a cutoff and set
\be
\label{cutoffSUformula}
\delta H_A = {i \epsilon \over 2} \int_{-\Lambda}^\Lambda {ds \over 1 + \cosh s} \left((G_A \big\vert_{s - i \pi} + \widetilde{G_{\bar{A}}} \big\vert_s) - (G_A \big\vert_{s + i\pi}
+ \widetilde{G_{\bar{A}}} \big\vert_s) \right)
\ee
We will see that the endpoint contribution survives as a finite effect as the cutoff is removed, $\Lambda \rightarrow \infty$.  This limit will be studied in
section \ref{sect:extract} where it gets rephrased as studying an equivalent limit $\delta \rightarrow 0$.

To interpret complex modular flow we proceed as follows.  Denote vacuum modular flow by
\be
\label{VacuumModularFlow}
J^{(n)}(x) \big\vert_s = e^{i \Hextvac s / 2 \pi} J^{(n)}(x) e^{-i \Hextvac s / 2 \pi}
\ee
Here $\Hextvac$ is the vacuum modular Hamiltonian, related to a Lorentz boost generator by $\Hextvac = 2 \pi K$ \cite{Hislop:1981uh,Witten:2018lha}.
By definition a field of modular weight $n$ satisfies
\be
J^{(n)}(x) \big\vert_s = e^{ns} J^{(n)}(e^s x^+, e^{-s} x^-, {\bf x}_\perp)
\ee
We'll also need the mirror or CPT conjugate operator\footnote{We're defining parity to act by reflecting the first spatial coordinate.}
\be
\label{CPT}
\widetilde{J^{(n)}}(x) = (-1)^n J^{(n)}(-x^+,-x^-,{\bf x}_\perp)
\ee
The S\'arosi--Ugajin formula involves complex modular flow, by modular time $s \pm i \pi$.  To make sense of this define
\be
\label{flowed}
G_A \big\vert_{s \pm i r} = \int_0^a dx^+ f(x^+) e^{n(s \pm ir)} J^{(n)}(e^{s \pm i r} x^+,-\theta e^{-(s \pm ir)} x^+,0)
\ee
Formally this is what one obtains by applying complex modular flow to (\ref{GAtheta}).  We also have
\be
\label{flowed2}
\widetilde{G_{\bar{A}}} \big\vert_s = \int_{-b}^0 dx^+ f(x^+) (-e^s)^n J^{(n)}(- e^s x^+,\theta e^{-s} x^+,0)
\ee
Our prescription for defining complex modular flow is to work with the operators (\ref{flowed}), (\ref{flowed2}) inside correlators and analytically continue $r \, : \, 0 \rightarrow \pi$.  An important fact is that, as we will discuss in section \ref{sect:GAcorrelators},
the combination which appears in the S\'arosi--Ugajin formula $G_A \big\vert_{s \pm i \pi} + \widetilde{G_{\bar{A}}} \big\vert_s$ has correlators which are well-behaved
at early and late modular times.  This will force us to work with the combination $G_A \big\vert_{s \pm i \pi} + \widetilde{G_{\bar{A}}} \big\vert_s$.  We believe this is a manifestation of the fact that the Hilbert space of the field theory cannot be tensor factored into ${\cal H}_A \otimes {\cal H}_{\bar{A}}$.

To decide what sort of correlators we should look at, note that when $r = 0$ both $G_A \big\vert_s$ and $\widetilde{G_{\bar{A}}} \big\vert_s$ only involve operators
inserted at
\be
\label{right}
x^+ > 0,\quad x^- < 0,\quad {\bf x}_\perp = 0
\ee
This means they have well-defined correlators with a spectator operator $J^{(n)}(y^+,y^-,{\bf y}_\perp)$ inserted in the left Rindler wedge
\be
\label{left1}
y^+ < 0, \quad y^- > 0
\ee
By inserting the spectator operator in this region we're placing it at spacelike separation and ensuring that correlators with $G_A \big\vert_{s}$ and
$\widetilde{G_{\bar{A}}} \big\vert_s$ are non-singular, at least for real values of $s$.

We'll make a stronger assumption, that to determine $\delta H_A$ it suffices to set ${\bf y}_\perp = 0$ and insert the spectator operators at
\be
\label{left2}
y^+ < 0, \quad y^- > 0, \quad {\bf y}_\perp = 0
\ee
To justify this we note that the spectator operator will not play much of a role in the calculations that follow, aside from giving us correlators that have a definite analytic structure to work with.  The spectator operator will be inserted then stripped off, and the results only depend on the
nature of the perturbation $G$.  From the S\'arosi--Ugajin formula we expect that, in any sensible interpretation of complex modular flow, $\delta H_A$ should be an operator constructed from $J^{(n)}$ in the region (\ref{right}).  For an operator $\delta H_A$ of this nature, it would seem spectator
operators in the region (\ref{left2}) should be adequate as a probe for determining $\delta H_A$.

A nice feature of setting ${\bf y}_\perp = 0$ is that for a given conformal dimension and modular weight correlators become universal, determined by the constraints of 2D global
conformal invariance.  So the only correlator we need to consider is
\be
\label{JJ}
\langle J^{(n)}(x^+,x^-,0) J^{(n)}(y^+,y^-,0) \rangle = {1 \over (x^+ - y^+)^{\Delta + n} (-x^- + y^-)^{\Delta - n}}
\ee
This resembles a correlator of primary fields in a 2D CFT with $h = {\Delta + n \over 2}$, $\bar{h} = {\Delta - n \over 2}$.\footnote{Although it's simplest to think about primary fields, the correlator of a descendant with
itself has exactly the same form.  So our results apply to descendants as well.}  We've distributed the minus signs so the correlator is well-defined
in the region (\ref{right}), (\ref{left2}).
More generally we could consider non-zero ${\bf y}_\perp$, in which case correlators of primaries are no longer diagonal in modular weight.  But this generalization leads to the same result for $\delta H_A$
so it won't be necessary for our purposes.  We present some explicit calculations in support of this claim in appendix \ref{appendix:yperp}.

Finally the following property of light-ray operators \cite{Kravchuk:2018htv,Faulkner:2016mzt,Casini:2017roe,Cordova:2018ygx,Kologlu:2019bco,Huang:2020ycs,Belin:2020lsr} will play an important role.  These are operators of the form
\be
{\cal L}^k [J^{(n)}] = \int_{-\infty}^\infty dx^+ (x^+)^k J^{(n)}(x^+,x^-,{\bf x}_\perp) \qquad \hbox{\rm for $k = 0,1,2,\ldots$}
\ee
We'll refer to $k$ as the moment of the light-ray operator.  By a small extension of \cite{Kravchuk:2018htv} it can be shown that ${\cal L}^k [J^{(n)}]$ annihilates
the conformal vacuum both to the left and the right provided $k < \Delta + n - 1$.
\be
\label{annihilate}
{\cal L}^k [J^{(n)}] \vert 0 \rangle = 0 = \langle 0 \vert {\cal L}^k [J^{(n)}] \qquad \hbox{\rm for $k < \Delta + n - 1$}
\ee
The argument runs as follows.  Ref.\ \cite{Kravchuk:2018htv} inserts ${\cal L}^0[J^{(n)}]$ on the far left (far right) in a correlator with a string of local operators and
shows that as a function of $x^+$ the integrand is analytic and falls off as $1/(x^+)^{\Delta + n}$ in the lower (upper) half plane.  So the $x^+$ contour can be closed
and the integral vanishes, which means ${\cal L}^0[J^{(n)}]$ annihilates the vacuum to the left (right).  Inserting a factor of $(x^+)^k$ we see that the same argument
goes through and ${\cal L}^k [J^{(n)}]$ annihilates the vacuum as in (\ref{annihilate}) provided $k < \Delta + n - 1$.

A small further extension is possible and will be relevant for us: if a light-ray moment of a primary annihilates the vacuum then so does the light-ray moment of all its descendants.
\be
\label{annihilate2}
{\cal L}^k [P_{\mu_1} \cdots P_{\mu_l} J^{(n)}] \vert 0 \rangle = 0 = \langle 0 \vert {\cal L}^k [P_{\mu_1} \cdots P_{\mu_l} J^{(n)}] \qquad \hbox{\rm for $k < \Delta + n - 1$}
\ee
For insertions of the momentum operators $P_-$ and $P_\perp$ this is just the statement that the derivative of zero is zero.  Insertions of $P_+$ are related to lower moments
by integrating by parts, but those lower moments already annihilate the vacuum.

As a simple example that illustrates these properties, consider the $k^{th}$ moment of the 2-point function (\ref{JJ}).  With a Wightman $i \epsilon$ prescription this is
\be
\label{LightRayIntegral}
\langle {\cal L}^k [J^{(n)}] \, J^{(n)}(y) \rangle = \int_{-\infty}^\infty dx^+ {(x^+)^k \over (x^+ - y^+ - i\epsilon)^{\Delta + n} (-x^- + y^- + i \epsilon)^{\Delta - n}}
\ee
There is a branch cut from $x^+ = y^+ + i \epsilon$ to $x^+ = - \infty + i \epsilon$.  In the lower half plane the integrand is analytic and falls off like $1/(x^+)^{\Delta + n - k}$.
Provided $k < \Delta + n -1$ we can close the contour in the lower half plane and the integral vanishes.  Acting with $\partial_{x^-}$ doesn't change this behavior, so
we have
\be
\label{LightRayIntegral2}
\langle {\cal L}^k [\partial_-^l J^{(n)}] \, J^{(n)}(y) \rangle = 0 \qquad \hbox{\rm for $k < \Delta + n -1$ with $l = 0,1,2,\ldots$}
\ee
We'll make use of this vanishing property in section \ref{sect:extract}.  Although we won't need it, acting with $\partial_{x^+}$ improves the fall-off at large $x^+$, which
in fact leads to a wider range
\be
\langle {\cal L}^k [\partial_+^l J^{(n)}] \, J^{(n)}(y) \rangle = 0 \qquad \hbox{\rm for $k < \Delta + n + l -1$ with $l = 0,1,2,\ldots$}
\ee

\section{Correlators of $G_A$ and $\widetilde{G_{\bar{A}}}$\label{sect:GAcorrelators}}
Here we record properties of correlators involving $G_A$ and $\widetilde{G_{\bar{A}}}$ that, although not particularly interesting in their own right, will be important in the next section.  
For explicit calculations to illustrate these claims see appendix \ref{appendix:correlators}.

\noindent
\underline{\em Branch cuts} \\
We've been careful to place our spectator operator in a region where the correlator is well-defined at real modular time.
Singularities can arise, however, when we analytically continue to complex $s$.  Depending on the situation it will be convenient to work in terms of either $z = e^s$ or $w = 1/z = e^{-s}$.
For example we have (recall that we're defining complex modular flow by continuing $r: 0 \rightarrow \pi$ inside correlators)
\be
\label{GAJ}
\langle G_A \big\vert_{s \pm i r} J^{(n)}(y) \rangle = \left({e^{\pm i r} \over w} \right)^n \int_0^a dx^+ f(x^+) \langle J^{(n)}\big({e^{\pm i r} x^+ \over w},-{\theta w x^+ \over e^{\pm i r}},0\big) J^{(n)}(y) \rangle
\ee
As far as the correlator on the right is concerned, the perturbation acts on a segment which stretches from the origin to the point $\big({e^{\pm i r} a / w},-{\theta w a / e^{\pm i r}}\big)$ in the
$(x^+,x^-)$ plane.  As can be seen in the figure, there is a singularity when $e^{\pm i r} a / w= y^+$ and one end of the stretched interval becomes null separated from the spectator operator.

\smallskip
\centerline{\includegraphics{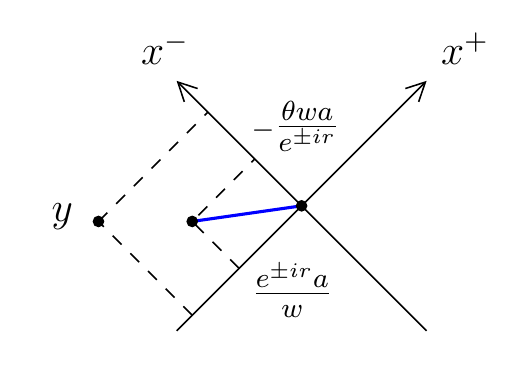}}

\noindent
The additional stretching as $w \rightarrow 0$ makes part of the stretched interval timelike separated from the spectator operator.  So generically
(\ref{GAJ}) has a branch cut which extends from $w = e^{\pm i r} a / y^+$ to $w = 0$.  There is also a light-cone singularity when $-{\theta w a \over e^{\pm i r}} = y^-$, so (\ref{GAJ}) also
has a branch cut that extends from $w = - {e^{\pm i r} y^- \over \theta a}$ to $w = - e^{\pm i r} \infty$.

\noindent
We also have
\be
\label{GAbartildeJ}
\langle \widetilde{G_{\bar{A}}} \big\vert_s J^{(n)}(y) \rangle = \left(- {1 \over w}\right)^n \int_{-b}^0 dx^+ f(x^+) \langle J^{(n)}\big(-{x^+ \over w},\theta w x^+,0\big) J^{(n)}(y) \rangle
\ee
By similar reasoning this is singular when $w = b / y^+$, with a branch cut that extends from $w = b / y^+$ to $w = 0$.  It is also singular when $w = - {y^- \over \theta b}$, with a branch cut that
extends from $w = - {y^- \over \theta b}$ to $w = - \infty$.

\noindent
\underline{\em Late modular time} \\
Now let's consider the behavior at small $w$.  Small $w$ corresponds to late modular time,
$s \rightarrow + \infty$, which stretches operators along the $x^+$ horizon.
It's important to note that (\ref{GAJ})
and (\ref{GAbartildeJ}) by themselves may diverge in this limit.  In fact both (\ref{GAJ})
and (\ref{GAbartildeJ}) behave $\sim 1/w^{n-1}$ as $w \rightarrow 0$, meaning they diverge in this limit
for $n \geq 2$.  However the S\'arosi--Ugajin formula only involves the combination (\ref{GAJ}) + (\ref{GAbartildeJ}).
For generic $r$ there is no cancellation and the sum also behaves $\sim 1/w^{n-1}$ as $w \rightarrow 0$.  But upon setting $r = \pi$ the sum can be represented by
\be
\label{GJ}
\langle \big(G_A \big\vert_{s \pm i \pi} + \widetilde{G_{\bar{A}}} \big\vert_s\big) J^{(n)}(y) \rangle = \left(- {1 \over w}\right)^n \int_{-b}^a dx^+ f(x^+) \langle J^{(n)}\big(-{x^+ \over w},\theta w x^+,0\big) J^{(n)}(y) \rangle
\ee
Since both limits of integration are non-zero this is finite and in fact vanishes $\sim w^\Delta$ as $w \rightarrow 0$, independent of the modular weight.

\noindent
\underline{\em Early modular time} \\
Finally we consider the behavior at large $w$ or early modular time.  This is the limit $s \rightarrow - \infty$ which stretches operators along the $x^-$ horizon.  The analysis parallels the previous situation.
At large $w$ we find that (\ref{GAJ}) and (\ref{GAbartildeJ}) both behave $\sim 1/w^{n+1}$, which means they diverge for $n \leq -2$.  For generic $r$ there is no cancellation and
the sum (\ref{GAJ}) + (\ref{GAbartildeJ}) also $\sim 1/w^{n+1}$ at large $w$.  However upon setting $r = \pi$ there is a cancellation: the sum (\ref{GJ})
is finite and in fact vanishes at large $w$.  It falls off $\sim 1/w^\Delta$, independent of the modular weight.

\section{Correlators of $\delta H_A$\label{sect:deltaHAcorrelators}}
To represent $\delta H_A$ we use the S\'arosi--Ugajin formula \cite{Sarosi:2017rsq} or the related expansion of \cite{Lashkari:2018tjh}, which for a perturbation of the form (\ref{G}) becomes
\be
\delta H_A = {i \epsilon \over 2} \int_{-\infty}^\infty {ds \over 1 + \cosh s} \left((G_A \big\vert_{s - i \pi} + \widetilde{G_{\bar{A}}} \big\vert_s) - (G_A \big\vert_{s + i\pi}
+ \widetilde{G_{\bar{A}}} \big\vert_s) \right)
\ee
In terms of $w = e^{-s}$ this is
\be
\label{SUw}
\delta H_A = i \epsilon \int_0^\infty {dw \over (w +1)^2} \left((G_A \big\vert_{s - i \pi} + \widetilde{G_{\bar{A}}} \big\vert_s) - (G_A \big\vert_{s + i\pi}
+ \widetilde{G_{\bar{A}}} \big\vert_s) \right)
\ee
We've kept terms grouped in this way -- despite what seems like an obvious cancellation -- since the grouped terms have well-behaved correlation
functions as $w \rightarrow 0$ and $w \rightarrow \infty$.  The need to keep operators grouped in this way is the manifestation inside correlators of the underlying fact that the
Hilbert space of a field theory can't be factored into a tensor product ${\cal H}_A \otimes {\cal H}_{\bar{A}}$.

To make sense of (\ref{SUw}) we take a step back and define
\be
\label{SUw2}
\delta H_A = i \epsilon \int_{0+}^{\infty-} {dw \over (w +1)^2} \left((G_A \big\vert_{s - i r} + \widetilde{G_{\bar{A}}} \big\vert_s) - (G_A \big\vert_{s + i r}
+ \widetilde{G_{\bar{A}}} \big\vert_s) \right)
\ee
Our procedure will be to work inside a correlator and compute $\langle \delta H_A J^{(n)}(y) \rangle$.
We define this by starting at spacelike separation where everything is well-defined and analytically continuing $r: 0 \rightarrow \pi$.
In (\ref{SUw2}) we've introduced a cutoff at early and late modular times, analogous to $\Lambda$ in (\ref{cutoffSUformula}), to deal with divergences as $w \rightarrow 0$ and $w \rightarrow \infty$.
Once we've continued to $r = \pi$ the divergences will cancel and we'll be able to study what happens as the cutoff is removed.

Now let's consider the analytic continuation of the correlator $\langle \delta H_A J^{(n)}(y) \rangle$ from $r = 0 \rightarrow \pi$.  To make the discussion concrete we assume $a,b > 0$ so that the perturbation straddles the
endpoint.  Then as we take $r: 0 \rightarrow \pi$ the first grouped term in (\ref{SUw2}) produces cuts from $e^{-ir} a / y^+$ to $0$ and from $-e^{-ir}y^-/\theta a$ to $-e^{-ir} \infty$ that rotate clockwise and hit the integration contour from above.  The
second grouped term in (\ref{SUw2}) produces cuts from $e^{ir} a / y^+$ to $0$ and from $-e^{ir}y^-/\theta a$ to $-e^{ir} \infty$ that rotate counterclockwise and hit the integration contour from below.
All these contributions can be combined into a pair of {\sf U} - shaped contours that wrap around the cuts.\footnote{We're assuming $-y^+ y^- > \theta a^2$, $-y^+ y^- > \theta b^2$ so the cuts are separated as shown.  This keeps the spectator operator
spacelike separated from the perturbation (\ref{Gthetadef}).}

\smallskip
\centerline{\includegraphics{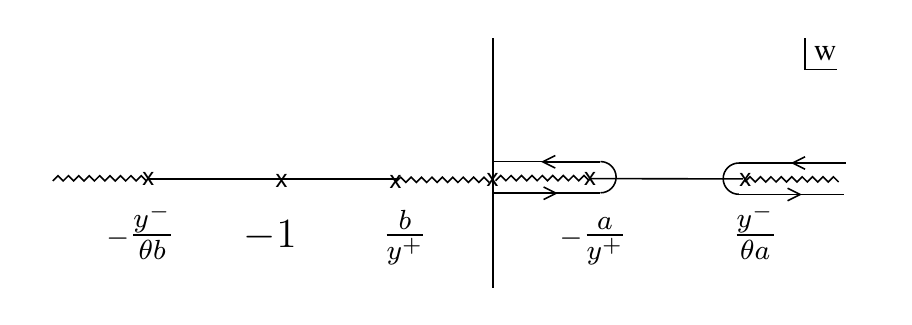}}

\noindent
Thanks to the cutoff on the integral over modular time in (\ref{SUw2}) we are able to keep the tips of the {\sf U}-shaped contours slightly separated.  So we're left with
\be
\label{deltaHAJ}
\langle \delta H_A J^{(n)}(y) \rangle = i \epsilon \int\limits_{\UU} {dw \over (w + 1)^2} \left(-{1 \over w}\right)^n \int_{-b}^a dx^+ f(x^+)
\langle J^{(n)}\big(-{x^+ \over w},\theta w x^+,0\big) J^{(n)}(y) \rangle
\ee
Although we derived this formula assuming $a,b > 0$ so that $G$ straddles the endpoint, in section \ref{sect:extract} we'll see that
this result is valid for all non-zero values of $a$ and $b$.

\section{Extracting $\delta H_A$\label{sect:extract}}
Our next job is to extract $\delta H_A$ from the correlator (\ref{deltaHAJ}).  There are three cases to consider: where the perturbation just acts on $A$, where the perturbation just
acts on $\bar{A}$, and where the perturbation straddles the endpoint.  In the first two cases we already know the answer, and in fact there is no need for perturbation theory: since
the unitary transformation $U$ factorizes as in footnote \ref{factorize} we have $H_A = U_A H_A^{(0)} U_A^\dagger$.  But as a warm-up exercise we show how this emerges from (\ref{deltaHAJ}) before turning to the more interesting case where the perturbation straddles the endpoint and the unitary transformation does not factorize.

\subsection{Perturbation acts on $A$\label{subsect:A}}
First let's consider the case where $a > -b > 0$ so that the perturbation just acts on $A$.  In this case $G_{\bar{A}}$ vanishes and the S\'arosi--Ugajin formula becomes
\bea
\nonumber
&& \delta H_A = {i \epsilon \over 2} \int_{-\infty}^\infty {ds \over 1 + \cosh s} \left(G_A \big\vert_{s - i \pi} - G_A \big\vert_{s + i\pi} \right) \\[5pt]
&& G_A \big\vert_{s \pm i r} = \left({e^{\pm i r} \over w} \right)^n \int_{-b}^a dx^+ f(x^+) J^{(n)}\big({e^{\pm i r} x^+ \over w},-{\theta w x^+ \over e^{\pm i r}},0\big)
\eea
By following the steps in section \ref{sect:deltaHAcorrelators} one sees that (\ref{deltaHAJ}) also applies to this situation.  The only real difference in the calculation
is that $\langle G_A \big\vert_{s \pm i r} J^{(n)}(y) \rangle$
now has additional cuts.  There are cuts from $w = e^{\pm i r} a / y^+$ to the origin and from $w = e^{\pm i r} (-b) / y^+$ to the origin.  There are also cuts from $w = - {e^{\pm i r} y^- / \theta a}$
to $- e^{\pm i r} \infty$ and from $w = - {e^{\pm i r} y^- / \theta (-b)}$ to $- e^{\pm i r} \infty$.  As $r : 0 \rightarrow \pi$ these cuts rotate and hit the integration contour from above
in the first term and from below in the second term.  So we end up with a pair of {\sf U} - shaped contours that wrap around the cuts.  This is correctly captured by (\ref{deltaHAJ}) which has cuts in the right places.

\smallskip
\centerline{\includegraphics{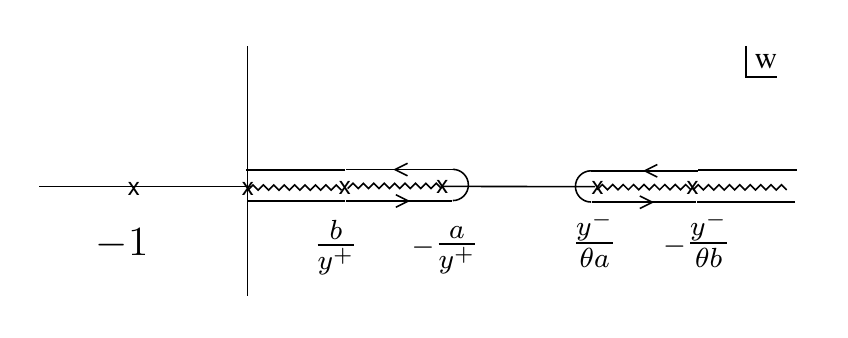}}

To extract $\delta H_A$ from (\ref{deltaHAJ}) we proceed as follows.  The correlator that appears in the integrand of (\ref{deltaHAJ}) is
\be
\label{GAthetaJ2}
\left(- {1 \over w}\right)^n \int_{-b}^a dx^+ f(x^+) \langle J^{(n)}\big(-{x^+ \over w},\theta w x^+,0\big) J^{(n)}(y) \rangle
\ee
Since both limits of integration are non-zero this vanishes $\sim w^\Delta$ as $w \rightarrow 0$ and vanishes $\sim 1/w^\Delta$ as $w \rightarrow \infty$.
So there is no problem closing the {\sf U}-shaped
contours in (\ref{deltaHAJ}) into contours that encircle the cuts.  The integrand vanishes at infinity, so at the cost of a minus sign instead of enclosing the
cuts we can take the contour to enclose the pole at $w = -1$.  Picking up the residue of the pole and stripping off the spectator operator $J^{(n)}(y)$ we obtain
\be
\delta H_A = 2 \pi \epsilon \left.{d \over dw}\right\vert_{w = -1} \left(- {1 \over w}\right)^n \int_{-b}^a dx^+ f(x^+) J^{(n)}\big(-{x^+ \over w},\theta w x^+,0\big)
\ee
Recognizing that $J^{(n)}$ is undergoing vacuum modular flow (\ref{VacuumModularFlow}) with $w = e^{-s}$ we can replace
\be
{d \over dw} \big(\cdot\big)= - {i \over 2 \pi w} \big[ \Hextvac, \cdot\, \big]
\ee
Setting $w = -1$ we obtain
\be
\delta H_A = i \epsilon \int_{-b}^a dx^+ f(x^+) \big[\Hextvac,J^{(n)}(x^+,-\theta x^+,0)\big]
\ee
or equivalently
\be
\delta H_A = i \epsilon \big[\Hextvac, G \big]
\ee
In this way we've recovered the expected result, that for a perturbation restricted to region $A$ the change in the
modular Hamiltonian is the commutator of $\Hextvac$ with the perturbation.

\subsection{Perturbation acts on $\bar{A}$\label{subsect:Abar}}
Next we consider the case $-b < a < 0$ where the perturbation just acts on $\bar{A}$.  In this case $G_A$ vanishes and
\be
\label{AbarPert}
\widetilde{G_{\bar{A}}} \big\vert_s = \left(- {1 \over w}\right)^n \int_{-b}^a dx^+ f(x^+) J^{(n)}\big(-{x^+ \over w},\theta w x^+,0\big)
\ee
Putting this in the S\'arosi--Ugajin formula (\ref{SUformula}) we rather trivially have $\delta H_A = 0$.  Alternatively one can say that
(\ref{deltaHAJ}) also applies to this situation, but that now the cuts are to the left of the origin.

\smallskip
\centerline{\includegraphics{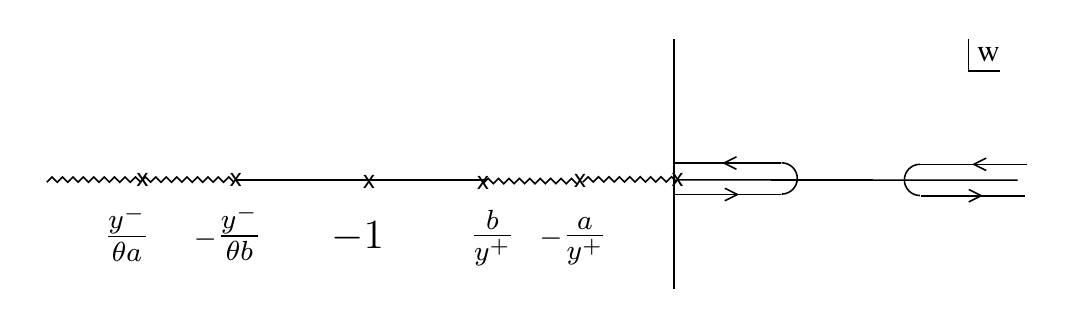}}

\noindent
Since both limits of integration in (\ref{AbarPert}) are non-zero we have
\be
\langle \widetilde{G_{\bar{A}}} \big\vert_s J^{(n)}(y) \rangle \sim
\left\lbrace\begin{array}{ll} w^\Delta & \qquad \hbox{\rm as $w \rightarrow 0$} \\
1/w^\Delta & \qquad \hbox{\rm as $w \rightarrow \infty$}\end{array}\right.
\ee
So there's no problem closing the {\sf U}-shaped contours and the integral vanishes since it encloses no singularities.
Again we get the expected result that $\delta H_A = 0$ for a perturbation that only acts on $\bar{A}$.

\subsection{Perturbation straddles the endpoint\label{subsect:straddle}}
Finally we consider the case $a,\,b > 0$ where the perturbation straddles the endpoint.  In this case it's useful to switch the order
of integration in (\ref{deltaHAJ}) and write (after stripping off the spectator operator)
\bea
\nonumber
&& \delta H_A = i \epsilon \int_{-b}^a dx^+ f(x^+) I(x^+) \\[5pt]
\label{I1}
&& I(x^+) = \int\limits_{\UU} {dw \over (w + 1)^2} \left(-{1 \over w}\right)^n J^{(n)}\big(-{x^+ \over w},\theta w x^+,0\big)
\eea
The claim, which we establish below by working inside correlators, is that for $n \geq 2$
\be
\label{I2}
I(x^+) = \Theta(x^+) \, {\cal C} +  \sum_{k = 0}^{n-2} {(-1)^k \over k!} \delta^{(k)}(x^+) \, {\cal E}_k
\ee
Here $\Theta(x^+)$ is a step function, $\delta^{(k)}(x^+) = \partial_+^k \delta(x^+)$ and the commutator and $k^{th}$ moment endpoint contributions are
\bea
\nonumber
&& {\cal C} = \big[\Hextvac, J^{(n)}(x^+,-\theta x^+,0)\big] \\[5pt]
\label{E}
&& {\cal E}_k = 2 \pi i \, \sum_{l = 0}^{\left\lfloor {n - 2 - k \over 2} \right\rfloor}  {(-\theta)^l \over l!} (n - k - 2l -1) \int_0^\infty dx^+ (x^+)^{k+l} \partial_-^l J^{(n)}(x^+,0,0)
\eea
This means that for $n \geq 2$
\bea
\label{endpoint1}
\delta H_A & = & - i \epsilon \int_0^a dx^+ f(x^+) \, \big[J^{(n)}(x^+,-\theta x^+,0), \Hextvac\big] \\
\nonumber
&& - 2 \pi \epsilon \sum_{k=0}^{n-2} {1 \over k!} \, f^{(k)}(0) \sum_{l = 0}^{\left\lfloor{n - 2 - k \over 2}\right\rfloor} {(-\theta)^l \over l!} (n-k-2l-1) \int_0^\infty dx^+ (x^+)^{k+l} \partial_-^l J^{(n)}(x^+,0,0)
\eea
Thus for $n \geq 2$, whenever the perturbation straddles the endpoint, the subregion modular Hamiltonian picks up an endpoint contribution on the $x^+$ horizon.
We'll see that for $n = -1, 0, 1$ the endpoint contribution vanishes while for $n \leq -2$ there is an analogous endpoint contribution on the $x^-$ horizon.

In the remainder of this section we specialize to $n \geq 0$ and seek to establish (\ref{I2}).  We proceed as follows.  Inserting (\ref{I1}) in a correlator with a spectator operator at
\be
\label{spacelike}
y^+ < 0, \quad y^- > 0, \quad {\bf y}_\perp = 0
\ee
we have
\be
\label{IJ}
\langle I(x^+) J^{(n)}(y) \rangle = \int\limits_{\UU} {dw \over (w + 1)^2} \left(-{1 \over w}\right)^n \langle J^{(n)}\big(-{x^+ \over w},\theta w x^+,0\big) J^{(n)}(y^+,y^-,0) \rangle
\ee
There is a light-cone singularity when $w = - x^+/y^+$ and the first argument of the two operators is the same.  Generically this produces a cut
from $w = - x^+ / y^+$ to $w = 0$.  There is another light-cone singularity when $w = y^- / \theta x^+$ and the second argument
of the two operators is the same.  Generically this produces a cut from $w = y^- / \theta x^+$ to $w = + \infty$ (if $x^+ > 0$) or to $w = - \infty$ (if $x^+ < 0$).
The situation for positive $x^+$ is shown in the figure.

\smallskip
\centerline{\includegraphics{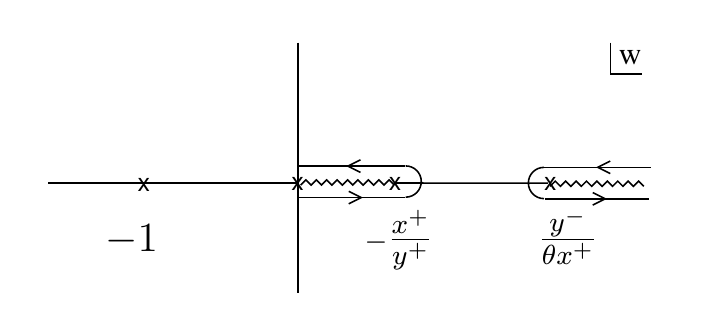}}

\noindent
Now let's see how (\ref{IJ}) depends on $x^+$.  To simplify the following discussion it's convenient to place the ends of the \leftU - shaped contour symmetrically at $w = \pm i \delta$.
The parameter $\delta \rightarrow 0^+$ will serve to regulate the distributions appearing in (\ref{I2}).  It's an avatar of the cutoff on modular time introduced in (\ref{cutoffSUformula}).

\begin{figure}
\centerline{\includegraphics[height=6cm]{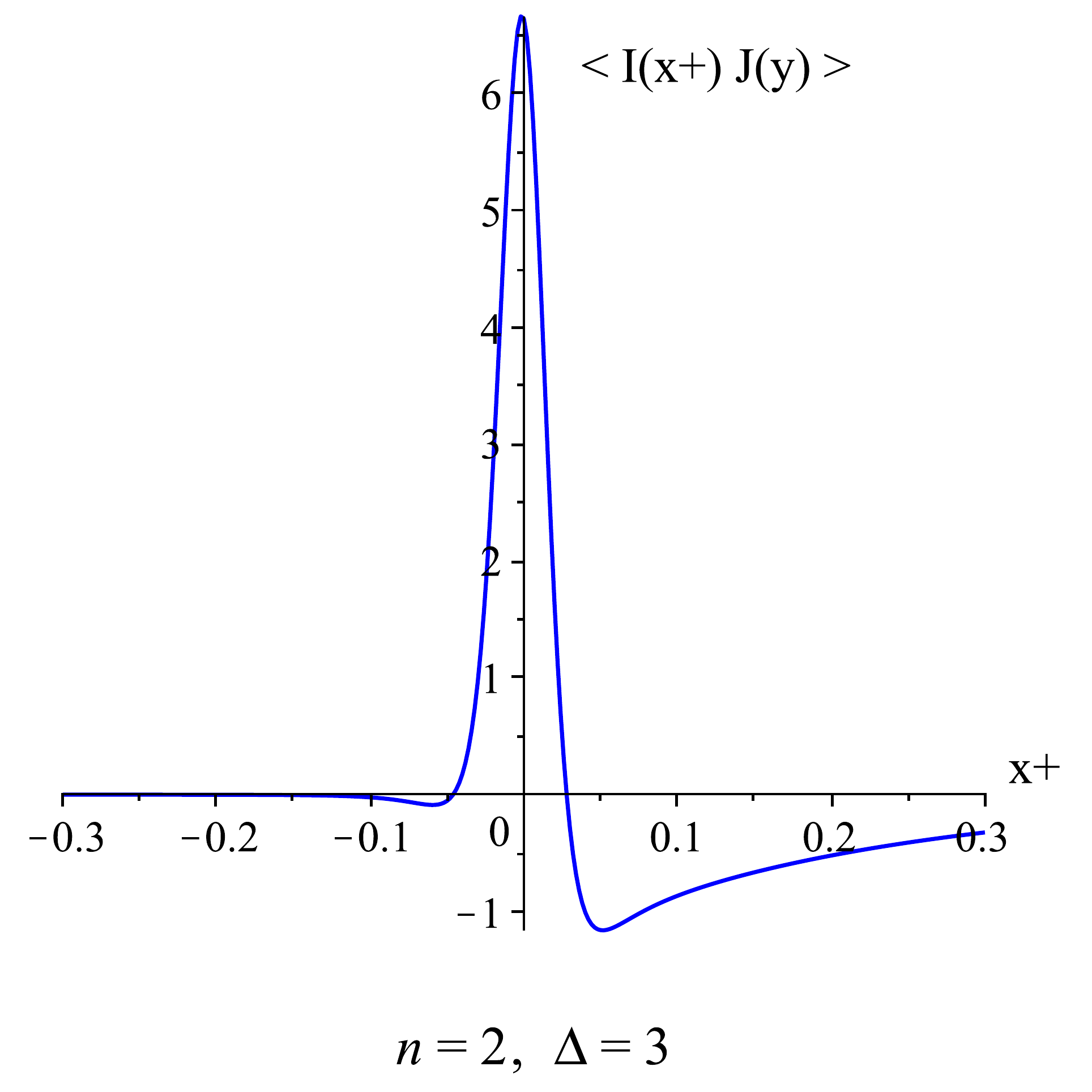} \includegraphics[height=6cm]{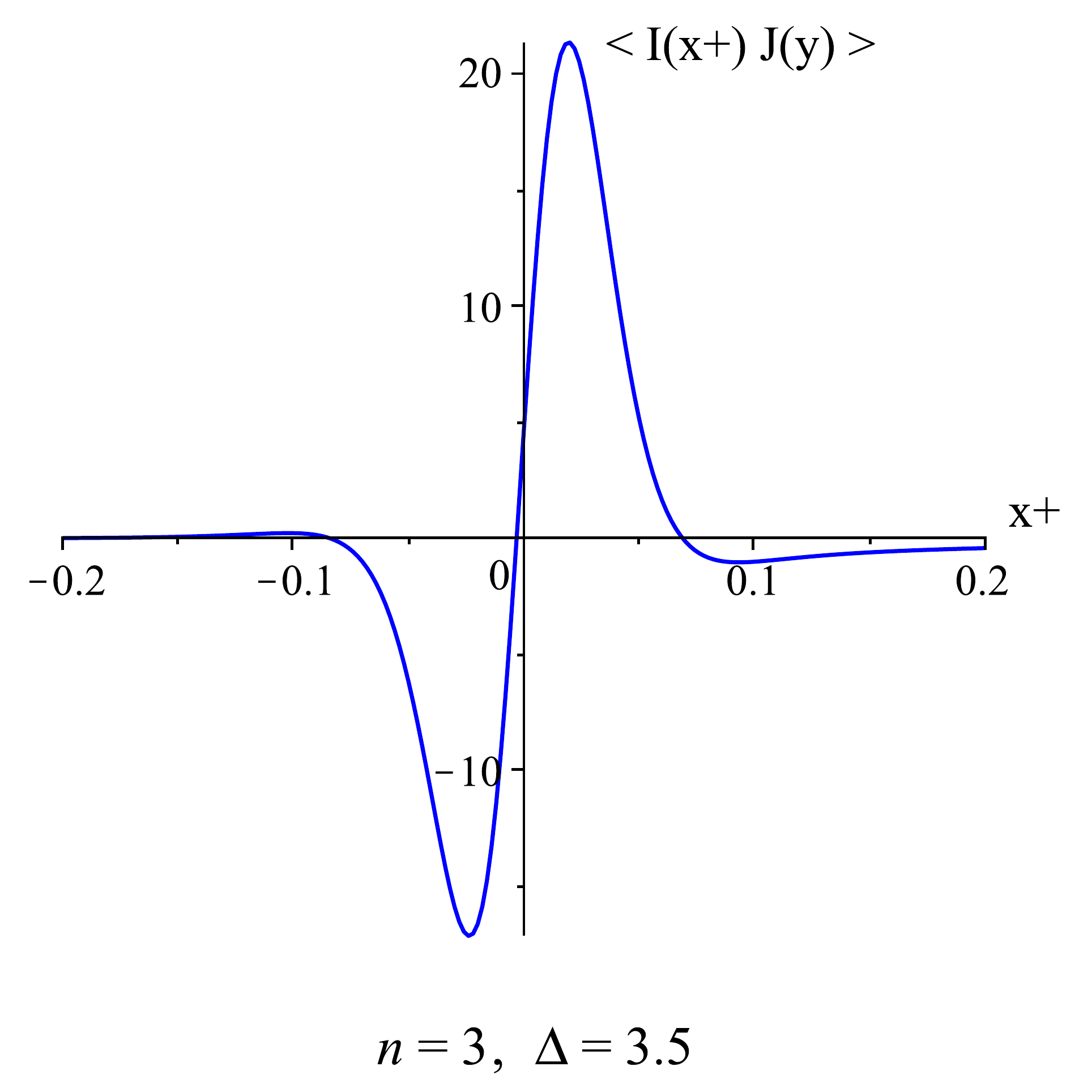} \includegraphics[height=6cm]{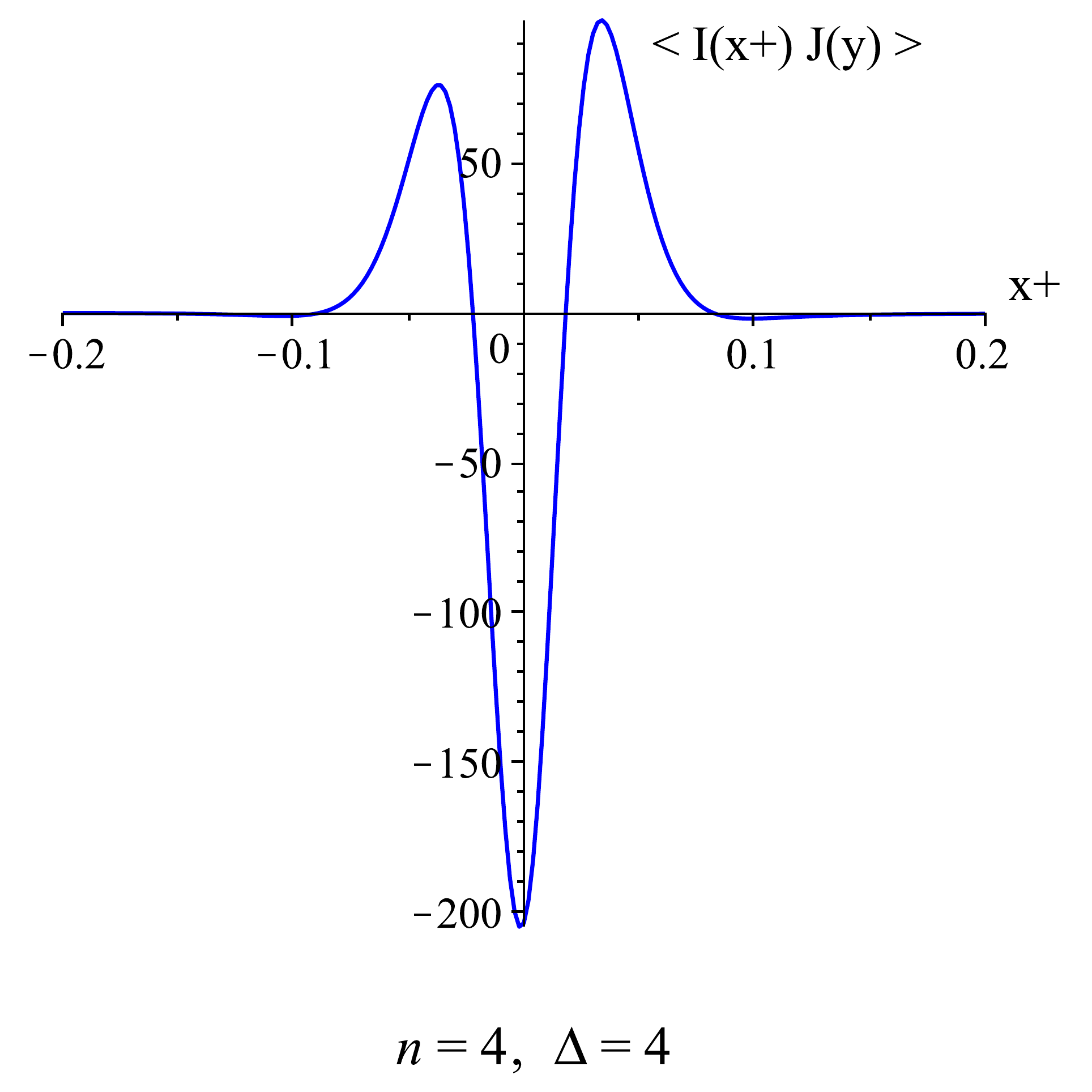}}
\caption{The imaginary part of $\langle I(x^+) J^{(n)}(y) \rangle$, defined in (\ref{I2d}), as a function of $x^+$ for various values of $n$ and $\Delta$.
Left panel: $n = 2$, $\Delta = 3$, $\delta = 0.03$.  Middle panel: $n = 3$, $\Delta = 3.5$, $\delta = 0.05$.  Right panel: $n = 4$, $\Delta = 4$, $\delta = 0.05$.
In all plots $y^+ = -1.5$, $y^- = 1.2$ and $\theta = 0$.\label{fig:Iplots}}
\end{figure}

The correlator (\ref{JJ}) leads to the explicit expression
\be
\label{I2d}
\langle I(x^+) J^{(n)}(y) \rangle = \int\limits_{\UU} {dw \over (w+1)^2} \left(-{1 \over w}\right)^n {1 \over \left(-{x^+ \over w\,\,} - y^+\right)^{\Delta + n}
\left(-\theta w x^+ + y^-\right)^{\Delta - n}}
\ee
Leaving aside the measure factor ${dw / (w + 1)^2}$, for $n \geq 0$ the integrand decays at infinity, with the behavior
\begin{equation}
\sim \left\lbrace\begin{array}{ll}1/w^n & \qquad \hbox{\rm if $x^+ = 0$} \\[3pt]
1/w^\Delta & \qquad \hbox{\rm if $x^+ \not= 0$}
\end{array}\right.
\end{equation}
So there is no problem closing the \rightU -shaped contour.  But as $w \rightarrow 0$ we have the behavior
\begin{equation}
\sim \left\lbrace\begin{array}{ll}1/w^n & \qquad \hbox{\rm if $x^+ = 0$} \\[3pt]
w^\Delta & \qquad \hbox{\rm if $x^+ \not= 0$}
\end{array}\right.
\end{equation}
As long as $x^+$ is non-zero there is no problem closing the \leftU -shaped contour.
But when $x^+ = 0$ there is a potential divergence and we need to be more careful.  To gain intuition about what to expect the correlator is plotted as a function of $x^+$ in Fig.\ \ref{fig:Iplots}.  One can clearly see the
appearance of singular behavior near $x^+ = 0$.  The singularity appears to involve $(-1)^k \delta^{(k)}(x^+)$ for $k = 0,\ldots,n-2$.

To study $I(x^+)$ analytically we first consider the possibility that $x^+$ is ``large.''  By this we just mean that $x^+$ is large compared to $\delta$, or more precisely that $\vert {x^+ \over \delta\,\,} \vert \gg \vert y^+ \vert$, where $\delta$ characterizes
the distance between the tips of the \leftU.  When $x^+$ is large in this sense we can neglect $y^+$ in (\ref{I2d}) near the tips of the \leftU.  Then the integrand vanishes $\sim w^\Delta$ near $w = 0$ and we can
bring the ends of the \leftU together to form a closed contour.  If $x^+ < 0$ the two closed contours that result (from closing \leftU and \rightU) can both be shrunk to a point and the integral
vanishes, while if $x^+ > 0$ we pick up a commutator with $\Hextvac$ as in section \ref{subsect:A}.  Thus as $\delta \rightarrow 0$ we find that
\be
\hbox{\rm large $x^+$ :} \quad I(x^+) = \Theta(x^+) \, \big[\Hextvac,J^{(n)}(x^+,-\theta x^+,0)\big]
\ee
That is, large $x^+$ gives the commutator contribution in (\ref{I2}).

We also need to consider the possibility that $x^+$ is ``small,'' meaning $\vert {x^+ \over \delta\,\,} \vert \lesssim \vert y^+ \vert$.  In this case $I(x^+)$ can diverge
from the singular behavior of the integrand $\sim 1/w^n$ as $w \rightarrow 0$.\footnote{This is the behavior when $x^+ = 0$.  More generally the integrand has this behavior when
$x^+ \rightarrow 0$ at least as fast as $w$.}  Note that the leading behavior as $w \rightarrow 0$ is independent of $\theta$.  However there can be subleading singularities,
suppressed by powers of $\theta w x^+$.  To take these into account we expand (\ref{IJ}) in powers of $\theta$ and write
\be
\label{IJexpand}
\langle I(x^+) J^{(n)}(y) \rangle = \sum_{l = 0}^\infty {1 \over l!} (-\theta x^+)^l \int\limits_{\UU} {dw \over (w + 1)^2} \left(-{1 \over w}\right)^{n-l}
\langle \partial_-^l J^{(n)}\big(-{x^+ \over w},0,0\big) J^{(n)}(y^+,y^-,0) \rangle
\ee

As we saw in Fig.\ \ref{fig:Iplots}, at small $x^+$ we anticipate that $I(x^+)$ contains terms proportional to $\delta^{(k)}(x^+)$ for $k = 0,\ldots,n-2$.  
To test this, and to determine the coefficients, we integrate $I(x^+)$ against $(x^+)^k$ over an interval $-\beta < x^+ < \alpha$.  Here $\alpha$ and $\beta$ are small, fixed
positive quantities that we will later send to zero.  Although they should not be confused with $a$ and $b$, they enter in the calculation in the
same way.  So from the discussion in section \ref{sect:deltaHAcorrelators} we expect that the correlator
\be
\label{intI1}
\int_{-\beta}^\alpha dx^+ (x^+)^k \langle I(x^+) J^{(n)}(y) \rangle
\ee
has two terms, one with a cut from $-\alpha / y^+$ to the origin and the other with a cut from $\beta / y^+$ to the origin.  [See the figure above (\ref{deltaHAJ}).  Note that in the expansion (\ref{IJexpand})
the other two cuts recede to infinity.]
To evaluate (\ref{intI1}) it's useful to switch the order of integration and to make a change of variables $x^+ \rightarrow - w x^+$.  These steps give
\bea
\nonumber
&& \int_{-\beta}^\alpha dx^+ (x^+)^k \langle I(x^+) J^{(n)}(y) \rangle \\
\nonumber
&& = \sum_{l = 0}^\infty {(-\theta)^l \over l!} \int\limits_{\rotatebox{90}{\mbox{${\textstyle \sf U}$}}} {dw \over (w + 1)^2} \left(-{1 \over w}\right)^{n-l}
\int_{-\beta}^\alpha dx^+ (x^+)^{k+l} \langle \partial_-^l J^{(n)}\big(-{x^+ \over w},0,0\big) J^{(n)}(y^+,y^-,0) \rangle \\
\nonumber
&& = - \sum_{l = 0}^\infty {(-\theta)^l \over l!} \int\limits_{\rotatebox{90}{\mbox{${\textstyle \sf U}$}}} {dw \over (w + 1)^2} \left(-{1 \over w}\right)^{n-k-2l-1} \\
\label{intI2}
&& \hspace{3cm} \times \Big(\int_{-\alpha/w}^\infty - \int^{\infty}_{\beta/w}\Big) dx^+ (x^+)^{k+l} \langle \partial_-^l J^{(n)}\big(x^+,0,0\big) J^{(n)}(y^+,y^-,0) \rangle \qquad\quad
\eea
The two terms in (\ref{intI2}) separately behave $\sim w^{\Delta + l}$ as $w \rightarrow 0$, so there is no obstacle to closing the contour
around each term separately.  The first term has a cut from $w = -\alpha/y^+$ to the origin that's enclosed by the contour while the second term has a cut from $w = \beta/y^+$ to the origin
that's outside the contour.\footnote{As an illustration of these properties, one of the relevant integrals is
\be
\int_{-\alpha/w}^\infty dx^+ {(x^+)^{k+l} \over (x^+ - y^+)^{\Delta + n}}
\ee
This has a cut from $w = -\alpha/y^+$ to the origin and (as can be seen by considering $w \rightarrow 0^-$) falls off $\sim w^{\Delta + n -  k - l - 1}$ as $w \rightarrow 0$.}
So only the first term survives and we're left with
\be
\label{IJ2}
- \sum_{l = 0}^\infty {(-\theta)^l \over l!} \oint\limits_{\rm cut} {dw \over (w + 1)^2} \left(-{1 \over w}\right)^{n-k-2l-1} \int_{-\alpha/w}^\infty dx^+ (x^+)^{k+l}
\langle \partial_-^l J^{(n)}\big(x^+,0,0\big) J^{(n)}(y^+,y^-,0) \rangle
\ee
In (\ref{IJ2}) there's a pole at $w = -1$ and for $k +2 l > n - 1$ a pole at infinity.  Rather than encircling the cut we can think of the contour as encircling the poles.  Setting $w = 1/z$ and stripping off the spectator $J^{(n)}(y)$
we're left with
\be
- \sum_{l = 0}^\infty {(-\theta)^l \over l!} \oint\limits_{z = 0,-1} {dz \over (z + 1)^2} (-z)^{n-k-2l-1} \int_{-\alpha z}^\infty dx^+ (x^+)^{k+l} \partial_-^l J^{(n)}(x^+,0,0)
\ee
As $\alpha, \beta \rightarrow 0$ this has a finite limit which gives us the coefficient of ${(-1)^k \over k!} \delta^{(k)}(x^+)$ in $I(x^+)$.  So we find that
\be
\hbox{\rm small $x^+$ :} \quad I(x^+) = \sum_k {(-1)^k \over k!} \delta^{(k)}(x^+) \, {\cal E}_k
\ee
where the coefficient is
\be
{\cal E}_k = - \sum_{l = 0}^\infty {(-\theta)^l \over l!} \oint\limits_{z = 0,-1} {dz \over (z + 1)^2} (-z)^{n-k-2l-1} \int_0^\infty dx^+ (x^+)^{k+l} \partial_-^l J^{(n)}(x^+,0,0)
\ee
Evaluating the contour integral we find (assuming $n \geq 2$, otherwise the integral vanishes)
\be
{\cal E}_k =
\left\lbrace\begin{array}{ll}
2 \pi i \sum_{l = 0}^{\left\lfloor{n - 2 - k \over 2}\right\rfloor} {(-\theta)^l \over l!} (n-k-2l-1) \int_0^\infty dx^+ (x^+)^{k+l} \partial_-^l J^{(n)}(x^+,0,0) & \quad k = 0,1,\ldots,n-2 \\
0 & \quad \hbox{\rm otherwise}
\end{array}\right.
\ee
Thus small $x^+$ gives the endpoint contribution to $I(x^+)$ in (\ref{I2}).  As a contribution to $\delta H_A$ it gives the second line of (\ref{endpoint1}).
\be
\label{endpoint2}
\delta H_A^{\rm endpoint} =
- 2 \pi \epsilon \sum_{k=0}^{n-2} {1 \over k!} \, f^{(k)}(0) \sum_{l = 0}^{\left\lfloor{n - 2 - k \over 2}\right\rfloor} {(-\theta)^l \over l!} (n-k-2l-1) \int_0^\infty dx^+ (x^+)^{k+l} \partial_-^l J^{(n)}(x^+,0,0)
\ee

\section{Generalizations\label{sect:generalize}}
The results we've presented so far have several straightforward generalizations.  In this section we present results for the extended modular Hamiltonian and for negative modular weights.
We introduce dependence on the transverse coordinates, consider perturbations on null planes, and discuss the extension of the results to perturbations on curved surfaces.

\noindent
\underline{\em Extended modular Hamiltonian} \\
We've obtained an expression for the change in the subregion modular Hamiltonian $\delta H_A$.  But analogous results hold for the complementary region,
and by combining them we can obtain the change in the extended modular Hamiltonian $\delta \Hext = \delta H_A - \delta H_{\bar{A}}$.

To obtain $\delta H_{\bar{A}}$ we use the CPT transformation (\ref{CPT}).  Note that CPT takes $x^\pm \rightarrow - x^\pm$, so it exchanges $A$ with $\bar{A}$.
This means we can apply CPT to our state, use (\ref{endpoint2}) to obtain $\delta H_A$ for the transformed state, then apply CPT again to obtain
$\delta H_{\bar{A}}$ for the original state.  Applying CPT to $\vert \psi \rangle = \big(\identity - i \epsilon G\big) \rule[+.05\baselineskip]{0pt}{\baselineskip}\vert 0 \rangle$ gives a new
state $\vert \widetilde{\psi} \rangle = \big(\identity + i \epsilon \widetilde{G} \big) \vert 0 \rangle$.  Note the change in sign, due to the fact that CPT is anti-unitary.  We get $\delta H_A$ for the transformed state using (\ref{endpoint2}) with $G \rightarrow - \widetilde{G}$, then apply CPT to the result to obtain $\delta H_{\bar{A}}$
for the original state.\footnote{CPT exchanges the regions but also transforms the state, so it takes
\be
\delta H_{A \, \vert \widetilde{\psi} \rangle} \rightarrow \delta H_{\bar{A} \, \vert \psi \rangle} \qquad \delta H_{\bar{A} \, \vert \widetilde{\psi} \rangle} \rightarrow \delta H_{A \, \vert \psi \rangle}
\ee}
The change in the extended modular Hamiltonian can then be obtained from  $\delta \Hext = \delta H_A - \delta H_{\bar{A}}$.  Note that $\delta H_{\bar{A}}$ enters with a minus sign, which cancels
the sign change in the transformed state.

The net result of this procedure is that $\delta \Hext$ is given by extending the range of integration in (\ref{endpoint2}) to cover the full $x^+$ horizon.  Thus for $n \geq 2$ we have
\be
\delta \Hext^{\rm endpoint} =
- 2 \pi \epsilon \sum_{k=0}^{n-2} {1 \over k!} \, f^{(k)}(0) \sum_{l = 0}^{\left\lfloor{n - 2 - k \over 2}\right\rfloor} {(-\theta)^l \over l!} (n-k-2l-1) \int_{-\infty}^\infty dx^+ (x^+)^{k+l} \partial_-^l J^{(n)}(x^+,0,0)
\ee

\noindent
\underline{\em Negative modular weight} \\
In section \ref{sect:extract} we obtained the endpoint contribution $\delta H_A$ for positive $n$
as a light-ray moment on the $x^+$ horizon.  It's straightforward to see that for negative $n$ there's an analogous contribution on the $x^-$ horizon.  The easiest way
to obtain it is to note that a parity transformation
$x \rightarrow -x$ acts by exchanging $x^+ = t + x$ with $x^- = t - x$ and acts on the modular weight (the difference in the number of $x_+$ and $x_-$ indices)
by $n \rightarrow -n$.  Parity also exchanges $A$ with $\bar{A}$, so for modular weights
$n = -2,-3,\ldots$ there must be an endpoint contribution to $\delta H_{\bar{A}}$, the change in the modular Hamiltonian for the complement, on the $x^-$ horizon.

To obtain an expression for $\delta H_{\bar{A}}$, instead of using (\ref{Gthetadef}), it's convenient to write the perturbation in a hatted form where $x^-$ is used to parametrize the segment.
\be
\label{Ghatdef}
G = \int_{-\hat{b}}^{\hat{a}} dx^- \, \hat{f}(x^-) J^{(n)}(-\hat{\theta} x^-,x^-,0)
\ee
By adapting (\ref{endpoint2}) to this situation we can immediately write the endpoint contribution for $n = -2,-3,\ldots$ as
\be
\label{deltaHAbar}
\delta H_{\bar{A}}^{\rm endpoint} =
- 2 \pi \epsilon \sum_{k=0}^{\vert n \vert - 2} {1 \over k!} \, \hat{f}^{(k)}(0) \sum_{l = 0}^{\left\lfloor{\vert n \vert - 2 - k \over 2}\right\rfloor} {(-\hat{\theta})^l \over l!} (\vert n \vert - k - 2l - 1)
\int_0^\infty dx^- (x^-)^{k+l} \partial_+^l J^{(n)}(0,x^-,0)
\ee
For negative modular weights the perturbation is most naturally presented in the form (\ref{Ghatdef}).  But we can rewrite $G$ in the form (\ref{Gthetadef}), where $x^+$ is used to parametrize the segment, via the dictionary
\be
\hat{\theta} = {1 \over \theta} \qquad \hat{f}(x) = {1 \over \theta} f\big(-{x \over \theta}\big) \qquad \hat{a} = \theta b \qquad \hat{b} = \theta a
\ee
Then the endpoint contribution for negative modular weight becomes
\be
\label{deltaHAbar2}
\delta H_{\bar{A}}^{\rm endpoint} =
- 2 \pi \epsilon \sum_{k=0}^{\vert n \vert - 2} {1 \over k!} \, f^{(k)}(0) \sum_{l = 0}^{\left\lfloor{\vert n \vert - 2 - k \over 2}\right\rfloor} {(-1)^{k+l} (\vert n \vert - k - 2l - 1) \over l! \, \theta^{k+l+1}}
\int_0^\infty dx^- (x^-)^{k+l} \partial_+^l J^{(n)}(0,x^-,0)
\ee
To obtain the change in the extended modular Hamiltonian one simply extends the range of integration to $\int_{-\infty}^\infty dx^-$ and introduces an overall sign due to the definition
$\delta \Hext = \delta H_A - \delta H_{\bar{A}}$.
\be
\delta \Hext^{\rm endpoint} =
+ 2 \pi \epsilon \sum_{k=0}^{\vert n \vert - 2} {1 \over k!} \, f^{(k)}(0) \sum_{l = 0}^{\left\lfloor{\vert n \vert - 2 - k \over 2}\right\rfloor} {(-1)^{k+l} (\vert n \vert - k - 2l - 1) \over l! \, \theta^{k+l+1}}
\int_{-\infty}^\infty dx^- (x^-)^{k+l} \partial_+^l J^{(n)}(0,x^-,0)
\ee

Parity makes the endpoint contribution for negative modular weight easy to obtain, but one can also trace its origin in the calculation.  In contrast to positive modular weight,
where the endpoint contribution arises from behavior at late modular time, the endpoint contribution for negative modular weight arises from behavior at early modular time.  When
$n$ is negative and $x^+ = 0$ there is a
potential singularity at early modular time which can obstruct closing the \rightU - shaped contour in (\ref{I2d}) at $w = \infty$.

\noindent
\underline{\em Transverse coordinates} \\
Instead of restricting to a perturbation of the form (\ref{Gthetadef}), we can introduce dependence on the transverse coordinates.  If we perturb the vacuum by
\be
\label{TransverseG}
G = \int_{-b}^a dx^+ \int d^{d-2} x_\perp \, f(x^+,{\bf x}_\perp) J^{(n)}(x^+,-\theta x^+,{\bf x}_\perp)
\ee
then to first order the endpoint contribution to the extended modular Hamiltonian simply picks up an integral over the transverse coordinates.  Denoting $f^{(k)}(x^+,{\bf x}_\perp) = \partial_+^k f(x^+,{\bf x}_\perp)$
this means
\begin{itemize}
\item
The endpoint contribution vanishes for modular weights $n = -1,0,1$.
\item
For modular weights $n = 2,3,4,\ldots$ the endpoint contribution is a sum of light-ray moments of $J^{(n)}$ on the $x^+$ horizon.
\bea
\nonumber
\delta \Hext^{\rm endpoint} & = &
- 2 \pi \epsilon \int d^{d-2} x_\perp \sum_{k=0}^{n-2} {1 \over k!} \, f^{(k)}(0,{\bf x_\perp}) \sum_{l = 0}^{\left\lfloor{n - 2 - k \over 2}\right\rfloor} {(-\theta)^l \over l!} (n-k-2l-1) \\
\label{TransversePositive}
&& \hspace{1cm} \int_{-\infty}^\infty dx^+ (x^+)^{k+l} \partial_-^l J^{(n)}(x^+,0,{\bf x}_\perp)
\eea
\item
For modular weights $n = -2,-3,-4,\ldots$ the endpoint contribution is a sum of light-ray moments of $J^{(n)}$ on the $x^-$ horizon.
\bea
\nonumber
\delta \Hext^{\rm endpoint} & = &
+ 2 \pi \epsilon \int d^{d-2} x_\perp \sum_{k=0}^{\vert n \vert - 2} {1 \over k!} \, f^{(k)}(0,{\bf x_\perp}) \sum_{l = 0}^{\left\lfloor{\vert n \vert - 2 - k \over 2}\right\rfloor} {(-1)^{k+l} (\vert n \vert - k - 2l - 1) \over l! \, \theta^{k+l+1}} \\
\label{TransverseNegative}
&& \hspace{1cm} \int_{-\infty}^\infty dx^- (x^-)^{k+l} \partial_+^l J^{(n)}(0,x^-,{\bf x}_\perp)
\eea
\end{itemize}

\noindent
\underline{\em Perturbations on null planes} \\
Finally we consider perturbations that act along null planes.
Let's begin with the $x^+$ horizon.  To obtain a perturbation that acts along the $x^+$ horizon we can start with
the spacelike perturbation (\ref{TransverseG}) and send $\theta \rightarrow 0^+$.  If the modular weight is positive there is no problem sending $\theta \rightarrow 0+$
in (\ref{TransversePositive}).  Thus for the null perturbation
\be
G = \int_{-b}^a dx^+ \int d^{d-2} x_\perp \, f(x^+,{\bf x}_\perp) J^{(n)}(x^+,0,{\bf x}_\perp) \qquad \hbox{\rm with $n \geq 2$}
\ee
the endpoint contribution is given by retaining just the $l = 0$ terms in (\ref{TransversePositive}).
\be
\delta \Hext^{\rm endpoint} =
- 2 \pi \epsilon \int d^{d-2} x_\perp \, \sum_{k=0}^{n - 2} {n - k - 1 \over k!} f^{(k)}(0,{\bf x_\perp})
\int_{-\infty}^\infty dx^+ (x^+)^k J^{(n)}(x^+,0,{\bf x}_\perp)
\ee
But for negative modular weight the limit $\theta \rightarrow 0+$ in (\ref{TransverseNegative}) appears singular.  It seems the modular Hamiltonian for a null perturbation with
negative modular weight is not well-defined.

To gain intuition as to why this might be the case it's helpful to boost to a frame where the perturbation acts on a fixed-time surface.  Consider the following perturbation, defined to act
on the slice $t = 0$.
\be
\label{Gt=0}
G_0 = \int_{-b}^a d\sigma \int d^{d-2} x_\perp \, f(\sigma,{\bf x}_\perp) \, \theta^{n \over 2} J^{(n)}\big(\sqrt{\theta} \, \sigma,-\sqrt{\theta} \, \sigma,{\bf x}_\perp\big)
\ee
As follows from (\ref{boost}), under a boost with $e^s = {1 \over \sqrt{\theta}}$ this turns into the perturbation of interest (\ref{TransverseG}).  But the form (\ref{Gt=0}) makes the limit $\theta \rightarrow 0$ easier to
understand.  For any modular weight as $\theta \rightarrow 0$ the perturbation becomes localized at $t = x = 0$.  If $n$ is positive the
coefficient in front of the
perturbation goes to zero and it's reasonable to expect the modular Hamiltonian to have a well-defined limit.\footnote{One might guess that $\delta \Hext$ would vanish as $\theta
\rightarrow 0$, on the grounds that in this limit $G_0 \rightarrow 0$, but that guess would be wrong.}  But if $n$ is negative the coefficient in front of the perturbation
diverges as $\theta \rightarrow 0$ and one might expect the modular Hamiltonian to diverge as well.\footnote{This guess does seem to be correct, or at least it's borne out by the $\theta \rightarrow 0$ divergences in (\ref{TransverseNegative}).}

We can likewise study perturbations which act along the $x^-$ horizon provided the modular weight is negative.  This corresponds to sending $\theta \rightarrow \infty$
in \ref{Gthetadef}, or better to
sending $\hat{\theta} \rightarrow 0$ in (\ref{Ghatdef}).  For a null perturbation
\be
G = \int_{-\hat{b}}^{\hat{a}} dx^- \int d^{d-2} x_\perp \, \hat{f}(x^-,{\bf x}_\perp) J^{(n)}(0,x^-,{\bf x}_\perp)  \qquad \hbox{\rm with $n \leq -2$}
\ee
the endpoint contribution to the extended modular Hamiltonian is a straightforward generalization of the (negative of the) $l = 0$ terms in (\ref{deltaHAbar}).
\be
\delta \Hext^{\rm endpoint} =
2 \pi \epsilon \int d^{d-2} x_\perp \, \sum_{k=0}^{\vert n \vert - 2} {\vert n \vert - k - 1 \over k!} \hat{f}^{(k)}(0,{\bf x_\perp})
\int_{-\infty}^\infty dx^- (x^-)^k J^{(n)}(0,x^-,{\bf x}_\perp)
\ee

\noindent
\underline{\em Perturbations on curved surfaces} \\
It's possible to generalize these results to perturbations that act on a curved spacelike surface.  For simplicity we introduce the perturbation at ${\bf x}_\perp = 0$ and
define the surface by setting $x^- = g(x^+)$ where $g(x^+)$ is a smooth function that vanishes at
$x^+ = 0$.  Thus in place of (\ref{Gthetadef}) we consider the perturbation
\be
\label{Gcurved}
G = \int_{-b}^a dx^+ f(x^+) J^{(n)}(x^+,g(x^+),0)
\ee
The tilted segment we worked with previously corresponds to the choice $g(x^+) = - \theta x^+$.  We'll discuss the subregion modular Hamiltonian $\delta H_A$ for this
perturbation, assuming positive $n$.  Other cases, such as negative modular weights and the extended modular Hamiltonian, may be treated in a similar manner.

To evaluate the endpoint contribution for this perturbation, note that the expansion which replaces (\ref{IJexpand}) is
\be
\langle I(x^+) J^{(n)}(y) \rangle = \sum_{l = 0}^\infty {1 \over l!} (g(x^+))^l \langle I_l(x^+) J^{(n)}(y) \rangle
\ee
where
\be
\label{IlJ}
\langle I_l(x^+) J^{(n)}(y) \rangle = \int\limits_{\UU} {dw \over (w + 1)^2} \left(-{1 \over w}\right)^{n-l}
\langle \partial_-^l J^{(n)}\big(-{x^+ \over w},0,0\big) J^{(n)}(y^+,y^-,0) \rangle
\ee
As in section \ref{subsect:straddle}, one can probe for terms proportional to $\delta^{(k)}(x^+)$ in $I_l(x^+)$ by integrating (\ref{IlJ}) against $(x^+)^k$.
For $n \geq 2$ this leads to the endpoint contribution for the subregion modular Hamiltonian
\be
\label{endpoint3}
\delta H_A^{\rm endpoint} = - 2 \pi \epsilon \sum_{k = 0}^{n - 2} \sum_{l = 0}^{n - 2 - k} {n - k - l - 1 \over k! \, l!} \left[ \left.{d^k \over dx^+{}^k}\right\vert_{x^+ = 0} f(x^+) (g(x^+))^l \right]
\int_0^\infty dx^+ (x^+)^k \partial_-^l J^{(n)}(x^+,0,0)
\ee
Upon setting $g(x^+) = - \theta x^+$ and re-arranging the sums this agrees with (\ref{endpoint2}).  The form (\ref{endpoint3}) makes the structure of the endpoint contribution more transparent;
we see that in general it involves a sum over moments $k$ and descendant levels $l$ bounded by $k + l \leq n - 2$.  But the form (\ref{endpoint2}) is more explicit for the case of perturbations
that act on a planar surface.

It's straightforward to extend this result to obtain the endpoint contribution to the extended modular Hamiltonian for a general curved surface.  We include dependence on the transverse coordinates
by introducing a function $g(x^+,{\bf x}_\perp)$ that vanishes at $x^+ = 0$.  Then for the perturbation
\be
\label{TransverseCurvedG}
G = \int_{-b}^a dx^+ \int d^{d-2} x_\perp \, f(x^+,{\bf x}_\perp) J^{(n)}(x^+,g(x^+,{\bf x}_\perp),{\bf x}_\perp)
\ee
the endpoint contribution for $n \geq 2$ is
\bea
\nonumber
\delta \Hext^{\rm endpoint} & = & - 2 \pi \epsilon \sum_{k = 0}^{n - 2} \sum_{l = 0}^{n - 2 - k} {n - k - l - 1 \over k! \, l!} \int d^{d-2} x_\perp \,
\left[ \left.{d^k \over dx^+{}^k}\right\vert_{x^+ = 0} f(x^+,{\bf x}_\perp) (g(x^+,{\bf x}_\perp))^l \right] \\
\label{endpoint4}
& & \hspace{0.5cm} \int_{-\infty}^\infty dx^+ (x^+)^k \partial_-^l J^{(n)}(x^+,0,{\bf x}_\perp)
\eea
The analogous result for negative modular weight can be obtained by exchanging $x^+ \leftrightarrow x^-$, replacing $n$ with $\vert n \vert$, and making an overall change of sign.

\section{Conclusions\label{sect:conclusions}}
We've seen that for small perturbations of the ground state the modular Hamiltonian for a division into half-spaces $\lbrace x < 0 \rbrace \cup \lbrace x > 0 \rbrace$ gets
a non-trivial endpoint contribution, arising whenever the perturbation can't be factored into a tensor product of unitary operators
on $A$ and $\bar{A}$.  We were able to capture the endpoint contribution by working inside correlation functions, where the lack of tensor factorization manifests itself
as potential divergences at early and late modular times.  With a cutoff on modular time in place the calculation becomes well-defined, and when the cutoff is removed
a finite endpoint contribution survives.  In this sense the endpoint contribution can be regarded as an anomaly in tensor factorization.

Extending our previous results \cite{Kabat:2020oic} we found that the endpoint contribution to the modular Hamiltonian involves light-ray moments of the form
\be
{\cal L}^{k} \, [\partial_-^l J^{(n)}]  = \int_{-\infty}^\infty dx^+ \, (x^+)^{k} \partial_-^l J^{(n)}(x^+,0,0)
\ee
for $k = 0,\ldots,n - 2 - l$.  This can be viewed as a generalization of the fact that the average null energy operator (the zeroth moment of the stress tensor)
\be
{\cal L}^0 \, [T_{++}] = \int_{-\infty}^\infty dx^+ \, T_{++}(x^+,0,0)
\ee
appears in the modular Hamiltonian under shape deformations \cite{Faulkner:2016mzt}; for the extension of this result to higher orders see \cite{Balakrishnan:2020lbp}.

Moments of light-ray operators satisfy some remarkable properties.  In particular they annihilate the conformal vacuum for moments satisfying $k < \Delta + n - 1$ \cite{Kravchuk:2018htv,Belin:2020lsr}.
It's worth noting that up to $k = n - 2$ this follows from the fact that these operators can appear in the modular Hamiltonian.  The modular Hamiltonian annihilates
the state, which to first order means
\be
\big(\Hextvac + \delta \Hext\big) \big(\identity - i \epsilon G \big) \vert 0 \rangle = 0
\ee
or
\be
\label{linearized}
\big(-i \epsilon \Hextvac G + \delta \Hext \big) \vert 0 \rangle = 0
\ee
Suppose the change in the modular Hamiltonian has the form $\delta \Hext = - i \epsilon [G,\Hextvac] + \delta \Hext^{\rm endpoint}$.  Substituting this in (\ref{linearized})
we see that $\delta \Hext^{\rm endpoint} \vert 0 \rangle = 0$.

In retrospect, in certain cases it's clear that endpoint contributions must be present whenever the perturbation doesn't vanish in a neighborhood of $x = 0$.  For example,
consider the vacuum state ``excited'' by applying a translation operator $e^{i \epsilon P_x}$.  Since the vacuum is invariant under a translation, the modular Hamiltonian (which is
uniquely determined by the state) must be invariant as well.
But the lowest-order modular Hamiltonian is a boost generator, $\Hextvac = 2 \pi K$, and this does {\em not} commute with translations.  So at first order in $\epsilon$
an endpoint contribution must be present to cancel the commutator term and make $\delta \Hext = 0$.  Note that the same argument applies to the modular conjugation operator
$J$ of Tomita - Takesaki theory, which at lowest order is identified with CPT: $J$ must receive endpoint contributions in perturbation theory.

Correlation functions of $G$ with a spectator operator have a well-defined analytic structure, discussed in section \ref{sect:GAcorrelators}.  We made heavy use of this structure
in obtaining the perturbed modular Hamiltonian.  But in addition to $G$ having well-defined correlators, one may wish to impose the condition that it produce a normalizeable
perturbation of the vacuum state: $\Vert \, G \vert 0 \rangle \, \Vert < \infty$.  For a recent discussion of normalizeability see section 5 of \cite{Bousso:2014uxa}.  The upshot is that the spacelike
perturbations we have considered are normalizeable in some but not all cases.  For example a conserved current in 2-D, or a scalar primary of dimension $\Delta < {d - 1 \over 2}$ in $d$
spacetime dimensions, produces a normalizeable perturbation.  Generically however some smearing in the time direction is required for normalizeability.  To understand the implications of this
we have begun exploring approaches for working with timelike perturbations.

Perturbation theory for modular operators is a large subject we have only begun to explore.  Various generalizations of our results are possible.  For example global conformal transformations
can be used to determine the endpoint contributions associated with spherical regions in a CFT.  It should be possible to extend the results to modular conjugation and to higher orders in perturbation theory.  We expect endpoint contributions to be present even in non-conformal theories provided they have a well-defined UV fixed point.
Finally note that we've adopted a particular method for calculating $\delta H_A$, based on the S\'arosi--Ugajin formula with a cutoff on modular time and a prescription for defining complex modular flow.  It would be interesting
to understand how the endpoint contribution arises from different methods with different regulators.
We hope to report on these developments in the near future.

\bigskip
\goodbreak
\centerline{\bf Acknowledgements}
\noindent
DK is supported by U.S.\ National Science Foundation grant PHY-1820734.  GL is supported in part by the Israel Science Foundation under grant 447/17.
PN acknowledges support from Israel Science Foundation grant 447/17 for the work in sections \ref{sect:intro} through \ref{sect:GAcorrelators} and appendix \ref{appendix:SUformula}
and from U.S.\ National Science Foundation grant PHY-1820734 for the work in sections \ref{sect:deltaHAcorrelators} through \ref{sect:conclusions} and appendix \ref{appendix:yperp}.

\appendix
\section{Path integral derivation of $\delta H_A$\label{appendix:SUformula}}
We present a Euclidean path integral derivation of the formula for $\delta H_A$ given in (\ref{SUformula}), which we interpreted in the bulk of the paper.  For an operator derivation of the
same result see appendix B of \cite{Kabat:2020oic}.  The steps here are formal.  They parallel the well-known Euclidean argument,
reviewed in section 5.2 of \cite{Witten:2018lha}, that for the ground state the reduced density matrix for the region $A = \lbrace x > 0 \rbrace$ is given by $\rho_A^{(0)} = e^{-2 \pi K}$.

We start from the state $\vert \psi \rangle = e^{-i \epsilon G} \vert 0 \rangle$.  If the unitary could be factored we would have
\be
\label{Euclideanpsi}
\vert \psi \rangle = \left(e^{-i \epsilon G_A} \otimes e^{-i \epsilon G_{\bar{A}}}\right)\vert 0 \rangle
\ee
This step isn't valid in general, or at the very least requires a regulator.  But we'll get sensible answers as long as we keep in mind that, as discussed below (\ref{flowed2}), the generators must always appear in the
combination $G_A + G_{\bar{A}}$.

The state (\ref{Euclideanpsi}) is produced by a Euclidean path integral on the half-space $\tau < 0$, with the operator $e^{-i \epsilon G_A} \otimes e^{-i \epsilon G_{\bar{A}}}$ inserted
on the $\tau = 0$ slice.\footnote{We're performing a Euclidean calculation, with the operator $G_A \otimes \identity_{\bar{A}} + \identity_A \otimes G_{\bar{A}}$ inserted at $\tau = 0$.
In Lorentzian signature this corresponds to a perturbation that acts at $t = 0$.  But this is adequate for analyzing the tilted case (\ref{Gthetadef}), since the operators appearing in (\ref{GAtheta}) and (\ref{GAbartheta})
can be evolved in time from $t = 0$.  Note that this evolution doesn't mix $A$ with $\bar{A}$.}  Likewise
\be
\langle \psi \vert = \langle 0 \vert \left(e^{i \epsilon G_A} \otimes e^{i \epsilon G_{\bar{A}}}\right)
\ee
is produced by a Euclidean path integral on the half-space $\tau > 0$ with operator insertions at $\tau = 0$.  Sewing these path integrals together along the complementary
region $\bar{A} = \lbrace x < 0 \rbrace$ and switching to angular evolution generated by $K$ we obtain an expression for the reduced density matrix of subregion $A$.
\bea
\nonumber
\rho_A = {\rm Tr}_{\bar{A}} \big( \vert \psi \rangle \langle \psi \vert \big)
& = & \raisebox{-9.2mm}{\includegraphics{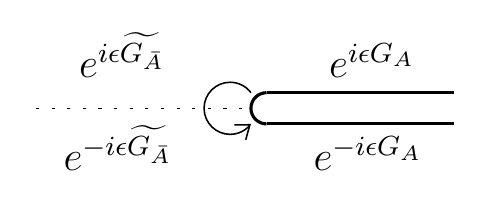}} \\
\label{rhoA}
& = & e^{-i \epsilon G_A} e^{-\pi K} e^{-i \epsilon \widetilde{G_{\bar{A}}}} e^{i \epsilon \widetilde{G_{\bar{A}}}} e^{-\pi K} e^{i \epsilon G_A}
\eea
Note that formally this is an operator on the Hilbert space for region $A$.  So by the time we've evolved through an angle $\pi$, rather than act with $G_{\bar{A}}$, we act with the CPT
conjugate operator $\widetilde{G_{\bar{A}}}$.  Intuitively this is because a Euclidean rotation by $\pi$ produces a CPT transformation, $J^{(n)}(x^+,x^-,{\bf x}_\perp) \rightarrow
(-1)^n J^{(n)}(-x^+,-x^-,{\bf x}_\perp)$.  The role of CPT can be seen more concretely in the operator formalism used in appendix B of \cite{Kabat:2020oic}.  Note that we've refrained
from making an obvious cancellation in (\ref{rhoA}) because we want to work with the combination $G_A + G_{\bar{A}}$.

From (\ref{rhoA}) the first-order change in the reduced density matrix is
\be
\delta \rho_A = - i \epsilon \left(G_A \rho_A^{(0)} + \big(\rho_A^{(0)}\big)^{1/2} \widetilde{G_{\bar{A}}} \big(\rho_A^{(0)}\big)^{1/2}\right)
+ i \epsilon \left(\rho_A^{(0)} G_A + \big(\rho_A^{(0)}\big)^{1/2} \widetilde{G_{\bar{A}}} \big(\rho_A^{(0)}\big)^{1/2}\right)
\ee
To compute $\delta H_A$ we use the expansion of the log developed in \cite{Sarosi:2017rsq,Lashkari:2018tjh}.  At first order this gives
\be
\delta H_A = - {1 \over 2} \int_{-\infty}^\infty {ds \over 1 + \cosh s} \big(\rho_A^{(0)}\big)^{-{1 \over 2} - {i s \over 2 \pi}} \delta\rho_A \big(\rho_A^{(0)}\big)^{- {1 \over 2} + {i s \over 2 \pi}}
\ee
Expressing this in terms of complex modular flow we obtain (\ref{SUformula}).
\be
\delta H_A = {i \epsilon \over 2} \int_{-\infty}^\infty {ds \over 1 + \cosh s} \left((G_A \big\vert_{s - i \pi} + \widetilde{G_{\bar{A}}} \big\vert_s) - (G_A \big\vert_{s + i\pi}
+ \widetilde{G_{\bar{A}}} \big\vert_s) \right)
\ee

\section{CFT correlators\label{appendix:correlators}}
We present some explicit calculations that exhibit our claims about the behavior of correlators.  We'll analyze the situation where $a,b > 0$ so that $G$ straddles the endpoint and return to
mention the other cases at the end.  
Then with the correlator (\ref{JJ}) we have
\bea
\nonumber
&& \langle G_A \vert_{s \pm i r} J^{(n)}(y) \rangle = \int_0^a dx^+ f(x^+) \left({e^{\pm i r} \over w}\right)^n {1 \over \left({e^{\pm i r} x^+ \over w} - y^+\right)^{\Delta + n}
\left({\theta w x^+ \over e^{\pm i r}} + y^-\right)^{\Delta - n}} \\
\label{GJexplicit}
&& \langle \widetilde{G_{\bar{A}}} \big\vert_s J^{(n)}(y) \rangle = \int_{-b}^0 dx^+ f(x^+) \left(-{1 \over w}\right)^n {1 \over \left(-{x^+ \over w} - y^+\right)^{\Delta + n}
\left(-\theta w x^+ + y^-\right)^{\Delta - n}}
\eea
These expressions have branch points when a singularity of the integrand collides with one of the limits of integration.  So $\langle G_A \vert_{s \pm i r} J^{(n)}(y) \rangle$
has a cut that runs from $w = e^{\pm ir} a / y^+$ to $w = 0$ and from $w = - e^{\pm i r} y^- / (\theta a)$ to $w = - e^{\pm i r} \infty$.  Likewise
$\langle \widetilde{G_{\bar{A}}} \big\vert_s J^{(n)}(y) \rangle$ has a cut from $w = b/y^+$ to $w = 0$ and from $w = - y^-/(\theta b)$ to $w = -\infty$.  These cuts were
also discussed in section \ref{sect:GAcorrelators}.

Now let's consider the behavior as $w \rightarrow 0$.  This behavior isn't sensitive to the function $f(x^+)$, so to capture it we set $f(x^+) = 1$.  Also to obtain
the leading behavior at small $w$ we set $\theta = 0$.  Then we have (here $C = - {1 \over (\Delta + n - 1) (-y^+)^{\Delta + n - 1} (y^-)^{\Delta - n}}$)
\bea
\label{appendixGAJ}
&&\langle G_A \big\vert_{s \pm i r} J^{(n)}(y) \rangle\strut_{\theta = 0} = C \left({e^{\pm ir} \over w}\right)^{n-1} \left({1 \over (- {e^{\pm i r} a \over w y^+} + 1)^{\Delta + n - 1}} - 1\right) \\
\label{appendixGAbartildeJ}
&&\langle \widetilde{G_{\bar{A}}} \big\vert_s J^{(n)}(y) \rangle\strut_{\theta = 0} = C \left(- {1 \over w}\right)^{n-1} \left(1 - {1 \over (-{b \over w y^+} + 1)^{\Delta + n - 1}} \right)
\eea
When $r = \pi$ these results can be combined to obtain
\be
\label{appendixGsumJ}
\langle \big(G_A \big\vert_{s \pm i \pi} + \widetilde{G_{\bar{A}}} \big\vert_s) J^{(n)}(y) \rangle\strut_{\theta = 0} = C \left(- {1 \over w}\right)^{n-1} \left({1 \over ({a \over w y^+} + 1)^{\Delta + n - 1}}
- {1 \over (-{b \over w y^+} + 1)^{\Delta + n - 1}} \right) \hspace{1cm}
\ee
These results illustrate our claims about behavior at small $w$.
For $n \geq 2$ both (\ref{appendixGAJ}) and (\ref{appendixGAbartildeJ}) diverge $\sim {1 / w^{n-1}}$ near $w = 0$.  But the divergence cancels in the sum (\ref{appendixGsumJ})
which remains finite and in fact vanishes $\sim w^\Delta$ near $w = 0$.

The procedure for studying behavior at large $w$ is very similar.  In (\ref{GJexplicit}), instead of setting $\theta = 0$, we would only retain the factors that do depend on $\theta$.
Then it's straightforward to perform the integrals explicitly and verify the claims about behavior at late modular time made in section \ref{sect:GAcorrelators}.

Finally let's see what happens when $G$ doesn't straddle the endpoint.  One possibility is $a > -b > 0$ where the perturbation only acts on $A$ and the relevant operator is
\be
G_A \big\vert_{s \pm i r} = \left({e^{\pm i r} \over w}\right)^n \int_{-b}^a dx^+ f(x^+) J^{(n)}\big({e^{\pm i r} x^+ \over w},-{\theta w x^+\over e^{\pm i r}},0\big)
\ee
Another possibility is $-b < a < 0$ where the perturbation only acts on $\bar{A}$ and the relevant operator is
\be
\widetilde{G_{\bar{A}}} \big\vert_s = \left(-{1 \over w}\right)^n \int_{-b}^a dx^+ f(x^+) J^{(n)}\big(-{x^+ \over w},\theta w x^+,0\big)
\ee
The behavior is the same in these two cases.  Thanks to the non-zero limits of integration these operators have correlators with a spectator operator $J^{(n)}(y)$ that vanish $\sim w^\Delta$
at small $w$ and fall off $\sim 1/w^\Delta$ at large $w$.

\section{Correlators at non-zero ${\bf y}_\perp$\label{appendix:yperp}}
In this appendix we consider certain analogs of the correlation functions studied in appendix \ref{appendix:correlators}, but with the spectator operator inserted at non-zero $\bold{y}_{\perp}$.
We will show that -- at least in the examples considered here -- the behavior of correlators is qualitatively similar to what we found in appendix \ref{appendix:correlators}.  This means the
calculations of sections \ref{sect:deltaHAcorrelators} and \ref{sect:extract} would go through in this more general setting and would lead to the same result for $\delta H_A$.

As noted in section \ref{sect:preliminaries}, at nonzero $\bold{y}_{\perp}$ the 2-point functions are no longer diagonal in modular weight. Furthermore we do not expect correlators to have universal
behavior, depending only on the modular weights of the operators involved; hence our notation will indicate the specific spin components of the operators.

We will consider two examples. In both examples we take the perturbation $G$ to be
\begin{equation}
G = \int_{-b}^{a} dx^{+} f(x^{+}) J_{+\dots+}{(x^{+},-\theta x^{+},0)}
\end{equation}
i.e.\ the light-ray of the all-plus component of a spin-$S$ primary operator $J$. In this section $S$ refers to the total number of indices of the current; note that our perturbation has modular weight
$n = S > 0$.  Also we will be interested in the behavior of correlation functions at small $w$ and large $w$. In these limits the details of the function $f(x^{+})$ do not matter, so we will set $f(x^{+}) = 1$
from now on.

In our first example we take the spectator operator to be $J_{+\dots+}{(y)}$ (the all-plus component of the same primary field), while in our second example we take the spectator operator to be the all-minus component $J_{-\dots-}{(y)}$. For both examples we use the 2-point function \cite{Giombi:2016ejx}
\begin{equation}\label{JJappendix}
\langle J_{S}{(x_{1}, \epsilon_{1})} J_{S}{(x_{2}, \epsilon_{2})} \rangle = \frac{C_{S}}{(x_{12}^{2})^{\Delta}} \left( \epsilon_{1} \cdot \epsilon_{2} - 2 \frac{\epsilon_{1} \cdot x_{12} \epsilon_{2} \cdot x_{12}}{x_{12}^{2}} \right)^{S}
\end{equation}
where $J_{S}{(x,\epsilon)} \equiv J_{\mu_{1}\mu_{2} \dots \mu_{S}} \epsilon^{\mu_{1}} \epsilon^{\mu_{2}} \dots \epsilon^{\mu_{S}}$ for some null polarization vector $\epsilon$, $x_{12} \equiv x_{1} - x_{2}$ and $C_{S}$
is a normalization factor.

\subsection{First example}
If we set both polarization vectors in (\ref{JJappendix}) to be $\partial_{+}$ we obtain the 2-point function
\begin{equation}
\langle J_{+\dots+}{(x^{+}, -\theta x^{+}, 0)} J_{+\dots+}{(y)} \rangle = \frac{C_{S}}{(-2)^{S}} \frac{(-\theta x^{+} - y^{-})^{2S}}{[(\theta x^{+} + y^{-})(x^{+} - y^{+}) + |\bold{y}_{\perp}|^{2}]^{\Delta+S}}
\end{equation}
Using this 2-point function, the correlators $\langle G_A \big\vert_{s \pm i r} J_{+\dots+}{(y)} \rangle$ and $\langle \widetilde{G_{\bar{A}}} \big\vert_s J_{+\dots+}{(y)} \rangle$ are
\begin{eqnarray}
\nonumber
&& \langle G_A \big\vert_{s \pm i r} J_{+\dots+}{(y)} \rangle =  \left( \frac{e^{\pm ir}}{w} \right)^{S} \frac{C_{S}}{(-2)^{S}} \int_{0}^{a} dx^{+} \frac{\left( -\frac{\theta w x^{+}}{e^{\pm ir}} - y^{-} \right)^{2S}}{\left[ \left( \frac{\theta w x^{+}}{e^{\pm ir}} + y^{-} \right) \left( \frac{e^{\pm i r}x^{+}}{w} - y^{+} \right) + |\bold{y}_{\perp}|^{2} \right]^{\Delta+S}} \\[5pt]
\nonumber
&& \langle \widetilde{G_{\bar{A}}} \big\vert_s J_{+\dots+}{(y)} \rangle = \left(-\frac{1}{w} \right)^{S} \frac{C_{S}}{(-2)^S} \int_{-b}^{0} dx^{+} \frac{\left( \theta w x^{+} - y^{-} \right)^{2S}}{\left[ \left(-\theta w x^{+} + y^{-} \right) \left( -\frac{x^{+}}{w} - y^{+} \right) + |\bold{y}_{\perp}|^{2} \right]^{\Delta+S}} \\
\label{BranchCuts}
\end{eqnarray}
As a function of $w$, these integrals produce branch points when a singularity of the integrand collides with an upper or lower limit of integration.  Thus
\begin{itemize}
\item $\langle G_A \big\vert_{s\pm i\pi} J_{+\dots+}{(y)} \rangle$ has a branch cut from $w = \frac{ay^{-}}{y^2}$ to $w = 0$.
\item $\langle \widetilde{G_{\bar{A}}} \big\vert_s J_{+\dots+}{(y)} \rangle$ has a branch cut from $w = -\frac{by^{-}}{y^2}$ to $w=0$.
\item $\langle G_A \big\vert_{s\pm ir} J_{+\dots+}{(y)} \rangle$ has a branch cut from $w = \frac{e^{\pm ir}y^{2}}{\theta y^{+}a}$ to $w = \infty$.
\item $\langle \widetilde{G_{\bar{A}}} \big\vert_s J_{+\dots+}{(y)} \rangle$ has a branch cut from $w = \frac{y^2}{\theta y^{+}b}$ to $w= \infty$.
\end{itemize}
These cuts are qualitatively similar to the ones found in appendix \ref{appendix:correlators}.

\paragraph{Small $w$ behavior.} To capture the small $w$ behavior we can set $\theta = 0$. We then obtain
\begin{eqnarray}
\label{corr1}
&& \langle G_A \big\vert_{s\pm ir} J_{+\dots+}{(y)} \rangle = \left( \frac{e^{\pm ir}}{w} \right)^{S-1} \frac{(-y^{-})^{2S-1}}{(\Delta+S-1)} \frac{C_{S}}{(-2)^S} \bigg[ \frac{1}{\left(\frac{e^{\pm ir}ay^{-}}{w} + y^{2} \right)^{\Delta+S-1}} - \frac{1}{(y^{2})^{\Delta+S-1}} \bigg] \hspace{1.5cm} \\
\label{corr2}
&& \langle \widetilde{G_{\bar{A}}} \big\vert_s J_{+\dots+}{(y)} \rangle = \left( -\frac{1}{w} \right)^{S-1} \frac{(-y^{-})^{2S-1}}{(\Delta+S-1)} \frac{C_{S}}{(-2)^S} \bigg[ \frac{1}{(y^{2})^{\Delta+S-1}} - \frac{1}{\left(\frac{b}{w}y^{-} + y^{2} \right)^{\Delta+S-1}} \bigg]
\end{eqnarray}
When $r=\pi$ these two correlators can be added to obtain (note the cancellation)
\be
\label{corr3}
\langle (G_A \big\vert_{s\pm i\pi} + \widetilde{G_{\bar{A}}} \big\vert_s) J_{+\dots+}{(y)} \rangle = \left(-\frac{1}{w} \right)^{S-1} \frac{(-y^{-})^{2S-1}}{(\Delta + S-1)} \frac{C_{S}}{(-2)^S} \bigg[ \frac{1}{\left( -\frac{ay^{-}}{w} + y^{2} \right)^{\Delta+S-1}} - \frac{1}{\left( \frac{b}{w}y^{-} + y^{2} \right)^{\Delta+S-1}} \bigg]
\ee
For $S \geq 2$ note that both (\ref{corr1}) and (\ref{corr2}) diverge $\sim 1/w^{S-1}$ as $w \rightarrow 0$.
But the divergence cancels in the sum (\ref{corr3}), which remains finite and in fact vanishes $\sim w^{\Delta}$ as $w \rightarrow 0$.
This is very similar to the behavior found in appendix \ref{appendix:correlators}.

\paragraph{Large $w$ behavior.} For large $w$ the correlators can be approximated as
\begin{eqnarray}
\nonumber
&& \langle G_A \big\vert_{s\pm ir} J_{+\dots+}{(y)} \rangle = \left( \frac{e^{\pm ir}}{w} \right)^{S} \frac{C_{S}}{(-2)^S} \int_{0}^{a} dx^{+} \frac{\left(\frac{-\theta w x^{+}}{e^{\pm ir}} - y^{-} \right)^{2S}}{\left( -\frac{\theta w x^{+}y^{+}}{e^{\pm ir}} + y^{2} \right)^{\Delta+S}} \\[5pt]
&& \langle \widetilde{G_{\bar{A}}} \big\vert_s J_{+\dots+}{(y)} \rangle = \left(-\frac{1}{w}\right)^{S} \frac{C_{S}}{(-2)^S} \int_{-b}^{0} dx^{+} \frac{(\theta w x^{+} - y^{-})^{2S}}{(\theta w x^{+}y^{+} + y^{2})^{\Delta+S}}
\end{eqnarray}
These integrals can be evaluated in terms of hypergeometric functions.
\begin{eqnarray}\label{corrA}
&& \langle G_A \big\vert_{s\pm ir} J_{+\dots+}{(y)} \rangle = \left( \frac{e^{\pm ir}}{w} \right)^{S+1} \frac{C_{S}}{(-2)^S (2S+1)} \frac{1}{\theta} \frac{1}{|\bold{y}_{\perp}|^{2(\Delta+S)}} \nonumber \\
&&\times \bigg[ \left( \frac{\theta wa}{e^{\pm i r}} + y^{-} \right)^{2S+1} {}_{2}F_{1}{\left( \Delta + S, 2S+1, 2S+2, \frac{y^{+}}{|\bold{y}_{\perp}|^2} \left(\frac{\theta w a}{e^{\pm ir}} + y^{-} \right) \right)} \nonumber \\
&& - (y^{-})^{2S+1} {}_{2}F_{1}{\left( \Delta + S, 2S+1, 2S+2, \frac{y^{+}y^{-}}{|\bold{y}_{\perp}|^2} \right)} \bigg]
\end{eqnarray}
\begin{eqnarray}\label{corrB}
&& \langle \widetilde{G_{\bar{A}}} \big\vert_s J_{+\dots+}{(y)} \rangle = \left(-\frac{1}{w} \right)^{S+1} \frac{C_{S}}{(-2)^S (2S+1)} \frac{1}{\theta} \frac{1}{|\bold{y}_{\perp}|^{2(\Delta+S)}} \nonumber \\
&&\times \bigg[ (y^{-})^{2S+1} {}_{2}F_{1}{\left( \Delta+S, 2S+1, 2S+2, \frac{y^{+}y^{-}}{|\bold{y}_{\perp}|^2} \right)} \nonumber \\
&& - (\theta w b + y^{-})^{2S+1} {}_{2}F_{1}{\left( \Delta+S, 2S+1, 2S+2, \frac{y^{+}}{|\bold{y}_{\perp}|^2} (\theta w b + y^{-}) \right)} \bigg]
\end{eqnarray}
When $r=\pi$ these correlators can be added to obtain (note the cancellation)
\begin{eqnarray}\label{corrC}
&& \langle (G_A \big\vert_{s\pm i\pi} + \widetilde{G_{\bar{A}}} \big\vert_s) J_{+\dots+}{(y)} \rangle = \left(-\frac{1}{w} \right)^{S+1} \frac{C_{S}}{(-2)^S (2S+1)} \frac{1}{\theta} \frac{1}{|\bold{y}_{\perp}|^{2(\Delta+S)}} \nonumber \\
&&\times \bigg[ \left(-\theta w a + y^{-} \right)^{2S+1} {}_{2}F_{1}{\left(\Delta + S, 2S+1, 2S+2, \frac{y^{+}}{|\bold{y}_{\perp}|^2} (-\theta w a + y^{-}) \right)} \nonumber \\
&&- (\theta w b + y^{-})^{2S+1} {}_{2}F_{1}{\left( \Delta+S, 2S+1, 2S+2, \frac{y^{+}}{|\bold{y}_{\perp}|^2} (\theta w b + y^{-}) \right)} \bigg]
\end{eqnarray}

To study the behavior at large $w$ we use the following identity, which follows from a $\zeta \rightarrow 1/\zeta$ transformation of
the hypergeometric function.
\begin{eqnarray}
\nonumber
&& {}_{2}F_{1}{(\Delta+S, 2S+1, 2S+2, \zeta)} = \frac{\Gamma(2S+2) \Gamma(\Delta-S-1)}{\Gamma(\Delta+S)} {1 \over (-\zeta)^{2S+1}} \\
\label{identity2}
&& \hspace{2cm} - \frac{2S+1}{(\Delta - S - 1)} {1 \over (-\zeta)^{\Delta+S}} \, {}_{2}F_{1}{\left( \Delta + S, \Delta - S - 1, \Delta - S, \frac{1}{\zeta} \right)} \hspace{1cm}
\end{eqnarray}
In all cases of interest $\Delta \geq S + 1$.\footnote{The unitarity bound requires $\Delta \geq S + d - 2$
and existence of a transverse direction requires $d \geq 3$.}  
Then the first term in (\ref{identity2}), which falls off $\sim 1/\zeta^{2S+1}$, gives the leading behavior of the hypergeometric function at large $\zeta$, while the second term gives subleading behavior $\sim 1/\zeta^{\Delta + S}$.

Using this result, the correlators (\ref{corrA}) and (\ref{corrB}) are seen to fall off $\sim 1 / w^{S+1}$ at large $w$.
As for the grouped correlator (\ref{corrC}), at large $w$ there is an exact cancellation of the leading behaviors of the two hypergeometric functions.  So it is the subleading behavior
(the second term in (\ref{identity2})) which determines the behavior at large $w$. We find that the grouped correlator (\ref{corrC}) falls off $\sim 1/w^{\Delta}$ at large $w$.  Again this is very
similar to the behavior found in appendix \ref{appendix:correlators}.

\subsection{Second example}
We now choose the first polarization vector in (\ref{JJappendix}) to be $\partial_{+}$ and the second polarization vector to be $\partial_{-}$. We then have the 2-point function
\begin{equation}
    \langle J_{+\dots+}{(x^{+},-\theta x^{+},0)} J_{-\dots-}{(y)} \rangle = \frac{C_{S}}{(-2)^S} \frac{|\bold{y}_{\perp}|^{2S}}{[(\theta x^{+} + y^{-})(x^{+} - y^{+}) + |\bold{y}_{\perp}|^{2} ]^{\Delta+S}}
\end{equation}
Using this 2-point function, the correlators $\langle G_A \big\vert_{s \pm ir} J_{-\dots-}{(y)} \rangle$ and $\langle \widetilde{G_{\bar{A}}} \big\vert_s J_{-\dots-}{(y)} \rangle$ are
\begin{eqnarray}
\nonumber
&& \langle G_A \big\vert_{s \pm ir} J_{-\dots-}{(y)} \rangle =
\left( \frac{e^{\pm ir}}{w} \right)^{S} \frac{C_{S}}{(-2)^S} |\bold{y}_{\perp}|^{2S} \int_{0}^{a} \frac{dx^{+}}{\left[ \left( \frac{\theta w x^{+}}{e^{\pm ir}} + y^{-} \right)\left( \frac{e^{\pm ir}x^{+}}{w} - y^{+} \right) + |\bold{y}_{\perp}|^{2} \right]^{\Delta+S}} \\[5pt]
\nonumber
&& \langle \widetilde{G_{\bar{A}}} \big\vert_s J_{-\dots-}{(y)} \rangle =
\left(-\frac{1}{w} \right)^{S} \frac{C_{S}}{(-2)^{S}} |\bold{y}_{\perp}|^{2S} \int_{-b}^{0} \frac{dx^{+}}{\left[ \left( -\theta w x^{+} + y^{-} \right) \left( -\frac{x^{+}}{w} - y^{+} \right) + |\bold{y}_{\perp}|^{2} \right]^{\Delta+S}} \\
\end{eqnarray}
Note that these correlators all vanish when $\bold{y}_{\perp} = 0$. This is as expected, since when ${\bf y}_\perp = 0$ the constraints of 2-dimensional conformal invariance require
that correlators be diagonal in modular weight.  Furthermore these correlators have exactly the branch cuts listed below (\ref{BranchCuts}).  This is also as expected, since the cuts
arise from light-cone singularities of the underlying 2-point function.

\paragraph{Small $w$ behavior.} To capture the behavior at small $w$ we can set $\theta = 0$. We then find (with $D_S = {C_S \over (-2)^S (\Delta + S - 1)}$)
\begin{eqnarray}
\label{corr4}
&& \langle G_A \big\vert_{s\pm ir} J_{-\dots-}{(y)} \rangle = -D_S\left( \frac{e^{\pm ir}}{w} \right)^{S-1} \frac{|\bold{y}_{\perp}|^{2S}}{y^{-}}
\left[ \frac{1}{\left( \frac{e^{\pm ir}ay^{-}}{w} + y^{2} \right)^{\Delta+S-1}} - \frac{1}{(y^{2})^{\Delta+S-1}} \right] \hspace{1.5cm} \\
\label{corr5}
&& \langle \widetilde{G_{\bar{A}}} \big\vert_s J_{-\dots-}{(y)} \rangle = -D_S\left(- \frac{1}{w} \right)^{S-1} \frac{|\bold{y}_{\perp}|^{2S}}{y^{-}}
\left[ \frac{1}{(y^{2})^{\Delta+S-1}} - \frac{1}{\left( \frac{by^{-}}{w} + y^{2} \right)^{\Delta+S-1}} \right]
\end{eqnarray}
When $r=\pi$ these correlators can be added to obtain (note the cancellation)
\be
\label{corr6}
\langle (G_A \big\vert_{s \pm ir} + \widetilde{G_{\bar{A}}} \big\vert_s) J_{-\dots-}{(y)} \rangle = -D_S\left(-\frac{1}{w}\right)^{S-1} \frac{|\bold{y}_{\perp}|^{2S}}{y^{-}}
\left[ \frac{1}{\left( -\frac{ay^{-}}{w} + y^{2} \right)^{\Delta+S-1}} - \frac{1}{\left( \frac{by^{-}}{w} + y^{2} \right)^{\Delta+S-1}} \right]
\ee
For $S \geq 2$ note that both (\ref{corr4}) and (\ref{corr5}) diverge $\sim 1/w^{S-1}$ as $w \rightarrow 0$.  But the divergence cancels in the sum (\ref{corr6}), which remains finite and in fact vanishes $\sim w^{\Delta}$ as $w \rightarrow 0$.  This is very
similar to the behavior found in appendix \ref{appendix:correlators}.

\paragraph{Large $w$ behavior.} At large $w$ the correlators are approximately given by
\begin{eqnarray}
\label{corr7}
&& \langle G_A \big\vert_{s \pm ir} J_{-\dots-}{(y)} \rangle = {D_S \over \theta} \left( \frac{e^{\pm ir}}{w} \right)^{S+1} \frac{|\bold{y}_{\perp}|^{2S}}{y^{+}}
\left[ \frac{1}{\left( -\frac{\theta w a y^{+}}{e^{\pm ir}} + y^{2} \right)^{\Delta+S-1}} - \frac{1}{(y^2)^{\Delta+S-1}} \right] \hspace{1.5cm} \\[5pt]
\label{corr8}
&&\langle \widetilde{G_{\bar{A}}} \big\vert_s J_{-\dots-}{(y)} \rangle = {D_S \over \theta} \left( -\frac{1}{w} \right)^{S+1} \frac{|\bold{y}_{\perp}|^{2S}}{y^{+}}
\left[ \frac{1}{(y^2)^{\Delta+S-1}} - \frac{1}{(-\theta wby^{+} + y^{2})^{\Delta+S-1}} \right]
\end{eqnarray}
When $r=\pi$ these correlators can be added to obtain (note the cancellation)
\be
\label{corr9}
\langle (G_A \big\vert_{s\pm ir} + \widetilde{G_{\bar{A}}} \big\vert_s) J_{-\dots-}{(y)} \rangle = {D_S \over \theta} \left(-\frac{1}{w}\right)^{S+1} \frac{|\bold{y}_{\perp}|^{2S}}{y^{+}}
\left[\frac{1}{(\theta w a y^{+} + y^{2})^{\Delta+S-1}} - \frac{1}{(-\theta w b y^{+} + y^{2})^{\Delta+S-1}} \right]
\ee
We see that the correlators (\ref{corr7}) and (\ref{corr8}) both fall off $\sim 1/w^{S+1}$ at large $w$, while the grouped correlator (\ref{corr9}) falls off $\sim 1/w^{\Delta+2S}$.
The grouped correlator has a faster fall-off than the $1/w^{\Delta}$ behavior found in appendix \ref{appendix:correlators}.  Fortunately faster decay still
allows the contour manipulations in section \ref{sect:extract} to go through,
so we would have obtained the same result for $\delta H_A$ even if we used a spectator operator 
$J_{-\dots-}(y)$ inserted at non-zero ${\bf y}_\perp$.


\providecommand{\href}[2]{#2}\begingroup\raggedright\endgroup

\end{document}